\documentclass[sigconf]{acmart}   

\usepackage{amsfonts}
\usepackage{algorithmic}
\usepackage{graphicx}
\usepackage{textcomp}
\usepackage{xcolor}
\def\BibTeX{{\rm B\kern-.05em{\sc i\kern-.025em b}\kern-.08em
    T\kern-.1667em\lower.7ex\hbox{E}\kern-.125emX}}
\usepackage{listings}
\usepackage{enumitem}
\usepackage{xspace}
\usepackage{tcolorbox}
\usepackage{subcaption}
\usepackage{booktabs}
\usepackage{url}

\definecolor{paleyellow}{rgb}{1, 1, 0.85}
\definecolor{lower}{rgb}{0.88235,0.7451,0.41569}
\definecolor{lower}{rgb}{0.94, 0.85, 0.6}
\definecolor{higher}{rgb}{0.5, 0.8, 0.75}
\definecolor{rowColor}{rgb}{0.937, 0.937, 0.937}
\definecolor{mauve}{rgb}{0.25,0,0.52}

\newcommand{\qualquote}[2]{\textit{\textcolor{mauve}{``#1'' (P#2)}}}
\newcommand{\qualquoteNoPID}[1]{\textit{\textcolor{mauve}{``#1''}}}

\newcommand{\summarytcolorbox}[1]{
\begin{tcolorbox}[colback=paleyellow, colframe=black, colbacktitle=white, coltitle=black, boxsep=0pt, top=3pt, bottom=3pt, left=6pt, right=6pt, before skip=3pt]
    \noindent #1
\end{tcolorbox}
}

\makeatletter
\patchcmd{\@makecaption}
  {\scshape}
  {}
  {}
  {}
\long\def\@makecaption#1#2{%
  \vskip\abovecaptionskip
  \small 
  \sbox\@tempboxa{#1. #2}%
  \ifdim \wd\@tempboxa >\hsize
    #1. #2\par
  \else
    \global \@minipagefalse
    \hb@xt@\hsize{\hfil\box\@tempboxa\hfil}%
  \fi
  \vskip\belowcaptionskip}
\makeatother

\newcommand{\quantPopSize}[0]{86\xspace}

\begin{document}

\newcommand{\pheading}[1]{\smallskip\noindent\textbf{#1}}


\title{Programmers Are Poor and Overconfident Judges of LLM-Generated Assertions}

\author{Zhanna Kaufman}
\email{zhannakaufma@umass.edu}
\orcid{0000-0000-0000-0000}
\affiliation{%
  \institution{University of Massachusetts Amherst}
  \city{Amherst}
  \state{MA}
  \country{USA}
}

\author{Yuriy Brun}
\email{brun@cs.umass.edu}
\affiliation{%
  \institution{University of Massachusetts Amherst}
  \city{Amherst}
  \state{MA}
  \country{USA}
}

\author{Adithya Murali}
\email{adithyamurali@cs.wisc.edu}
\affiliation{%
  \institution{University of Wisconsin–Madison}
  \city{Madison}
  \state{WI}
  \country{USA}
}

\author{Madeline Endres}
\email{mendres@umass.edu}
\affiliation{%
  \institution{University of Massachusetts Amherst}
  \city{Amherst}
  \state{MA}
  \country{USA}
}
\setcopyright{none}
\settopmatter{printacmref=false}
\renewcommand\footnotetextcopyrightpermission[1]{}


\begin{abstract}
Code comprehension and code review are already critically important software
engineering tasks, and the rising use of AI code generation tools is only
increasing that importance. Generative AI has the possibility of supporting
these activities, for example by augmenting code with assertions and 
natural-language explanations describing code behavior. However, 
little is known about
how effective such support may be. We conduct a controlled experiment with
\quantPopSize Python programmers and a follow-up think-aloud study to examine
developers' ability to assess the correctness and completeness of generated
assertions of varying quality, and to investigate how natural-language 
explanations influence these assessments.
While programmers can somewhat accurately judge correct assertions (74\% accuracy),
they perform poorly when shown incorrect assertions (49\% accuracy), despite 
reporting similar levels of confidence in both judgments. 
This difference in judgment accuracy is statistically significant ($p < 0.001$): the 
odds of a developer accurately judging a correct assertion was nearly three times higher 
than the odds of accurately judging an incorrect assertion (OR = 2.94). 
Surprisingly, natural-language explanations of assertions provided no overall 
benefit. Furthermore, low-quality explanations could impair specification assessment accuracy
($p=0.037$, $OR=0.58$) while simultaneously increasing developer confidence
($p = 0.005$, $3.99/5$ vs. $4.25/5$). 
Our findings suggest that, contrary to common assumptions, AI assistance may
not improve the reliability of code comprehension and review. 
More broadly, our findings highlight the importance of helping developers 
evaluate machine-generated reliability artifacts, in addition to generating them.

\end{abstract}

\maketitle

\section{Introduction}
\label{sec:intro}

\pagestyle{plain}
\thispagestyle{plain}

Generative AI tools have
transformed software engineering, and are now a routine part of virtually
every aspect of development including code generation~\cite{StackOverflow2025Survey}, 
architectural design~\cite{Jahic24}, debugging~\cite{StackOverflow2025Survey}, and
verification~\cite{First23fse}.
This makes code comprehension and code review, already important parts of the
software engineering lifecycle, even more important, as humans are asked to 
be part of the loop, reviewing, adapting, and adopting automatically
generated code~\cite{zhong2026human, pimenova2025good}. 
However, reviewing code is difficult, time-consuming,
and cognitively demanding~\cite{bacchelli2013expectations, baum2019associating}.
Programmers often struggle to understand non-self-authored code~\cite{Xia18},
including LLM-generated code~\cite{Vaithilingam22}, as such code may be
incorrect or misaligned with user intent~\cite{Zhong24, Fakhoury24}, which
can be difficult to fix in practice~\cite{Zi25}.
To reduce this burden, researchers have proposed shifting developer review
toward smaller reliability-oriented artifacts such as tests~\cite{lemieux2023codamosa, Fakhoury24},
specifications~\cite{zietsman2026specification, lahiri2026intent}, or assertions~\cite{Endres24}.
For example, Endres \emph{et al.} propose using LLMs to generate executable 
assertions that allow developers to review intended program behavior rather 
than the full implementation, reducing the amount of code that must be 
inspected while enabling run-time validation of generated code. 
Similarly, Lahiri envisions a future in which LLMs help formalize user 
intent through behavioral artifacts such as tests, assertions, and formal 
specifications~\cite{lahiri2026intent}.
By formalizing intended program behavior, such artifacts have the
potential to help developers determine if generated code captures their
intent and, in some cases, validate that intent at run time.

A key assumption underlying these approaches is that developers can
accurately evaluate machine-generated reliability artifacts. Yet this
assumption remains largely unexamined.
While a growing body of work studies developers' interactions and
understanding of LLM-generated code (see Hou 
\emph{et al.}~\cite{hou2024large}, Table 10), comparatively little empirical
evidence exists regarding how effectively developers evaluate generated
specifications, tests, or assertions.
This uncertainty extends to one of the most common mechanisms used to support
comprehension of such artifacts: natural-language explanations~\cite{Misra20,Hu22}. 
Modern LLMs routinely generate comments and explanations alongside code and
specifications.
However, like the code
itself, such explanations may be incorrect, incomplete, vague, or
irrelevant~\cite{Kang24,Cihan25}.
Consequently, it is unknown whether natural-language explanations
improve programmer judgment or instead introduce additional sources of error.

In this paper, we investigate two questions: 
%
(1)~How effective are programmers at judging the correctness and completeness
of LLM-generated assertions about code? and
(2)~How does the presence and quality of natural-language comments affect
these judgments?

We focus on postcondition assertions for several reasons.
First, postconditions possess well-defined formal semantics, allowing us to
rigorously characterize metrics such as correctness and completeness. Second,
LLMs are fairly proficient (but not yet perfect) at generating valid
assertions from documentation and have been explicitly proposed as part of a
human-in-the-loop software reliability workflow~\cite{Endres24}.
This allows us to study a close-to-ideal setting, in which LLMs are 
generally capable of generating useful assertions, but are not sufficiently 
reliable to eliminate the need for careful human review. 
%
Finally, postconditions can express rich 
behavioral properties while remaining
compact enough to support controlled experimental evaluation.

We conduct a controlled experiment with 86 Python programmers spanning a
range of professions and experiences.
We construct a corpus of Python functions paired with
LLM-generated postcondition assertions and natural-language comments,
extending datasets used in prior work on assertion
generation~\cite{Endres24,chen2021codex,evalplus}.
Participants were shown a function and postcondition, then asked to assess
whether the postcondition correctly captured the function's behavior and
how completely it did so.
To evaluate how explanation quality affects developers' reasoning about
generated reliability artifacts, we included four types of comments
that described the assertions: correct, incorrect, under-specified, and
over-specified.
We complemented the controlled experiment with a think-aloud study
with 10 developers.
This qualitative analysis allows us to characterize how developers reason
about generated assertions and explanations.

Our findings reveal a substantial asymmetry in developer reasoning.
Participants were much better at identifying correct 
assertions than identifying
incorrect ones (74\% vs.\ 49\% accuracy, $p < 0.001$, $OR = 2.94$),
despite reporting similarly high confidence in 
both cases (approximately $~4/5$ for both, $p > 0.1$).
Surprisingly, natural-language explanations led to no 
improvement in judgment accuracy ($p > 0.05$), and low-quality 
explanations could even reduce accuracy ($p = 0.037$, $OR = 0.58$) 
while simultaneously increasing confidence compared 
to no comment at all ($p = 0.005$, 4.25/5 vs. 3.99/5).

Simultaneously, developers' assessments of assertion completeness correlated
with an independent mutation-based measure of how many 
bugs it could detect (Spearman's $\rho = 0.18$, $p < 0.001$),
and participants rated correct assertions as significantly more complete
than incorrect ones ($p < 0.001$). Together, this 
suggests that developers can distinguish stronger specifications
from weaker ones even when they struggle to identify logical errors.
Providing insight into these findings, our think-aloud study 
suggests that developers frequently rely on direct
comparisons between assertion clauses and documentation, helping
explain why some incorrect assertions are difficult to detect. 
We make the following contributions:

\begin{itemize}[leftmargin=*]

  \item A controlled experiment with \quantPopSize Python programmers
  investigating how developers evaluate LLM-generated
  postconditions, a representative class of machine-generated
  specifications used in AI-assisted software reliability workflows.

  \item Evidence that developers are much better at recognizing
  correct assertions than incorrect ones 
  (74\% vs. 49\% accuracy, $p < 0.001$, $OR = 2.94$),
  despite reporting similarly high confidence in both cases
  (approximately 4/5, $p > 0.1$).

  \item An investigation of natural-language explanations, showing that
    they provide no overall improvement in developers' ability to evaluate
    assertions, while low-quality explanations can reduce accuracy
    ($p = 0.037$, $OR = 0.58$) compared to high-quality ones, and 
    simultaneously increase developer confidence ($p = 0.005$)
    compared to no comment at all.

  \item A think-aloud study with 10 developers identifying five recurring
    reasoning strategies for evaluating generated assertions and providing
    qualitative insight into why some classes of incorrect assertions may be
    systematically overlooked. 

\end{itemize}

We conclude by discussing implications for AI-assisted software
reliability, suggesting that future tools should move beyond generating
reliability artifacts to helping developers verify that they faithfully
capture intended program behavior.

\section{Motivating Example}
\label{sec:motivation}

Consider Juan, a developer who is asked to write and test the function
{\fontsize{8.9}{10}\selectfont\texttt{has\_close\_elements(numbers: List[float], threshold: float)}}
 (this
example comes from the HumanEval~\cite{chen2021codex} dataset). Juan is given
a natural-language docstring description of the desired functionality:

\begin{lstlisting}[basicstyle=\ttfamily\footnotesize, breakindent=\fontcharwd\font`a]
"Check if in given list of numbers, are any two numbers closer to each other than given threshold."
\end{lstlisting}

Juan decides to use an LLM to generate and test the code. 
%
%
In generating unit tests, the biggest challenge lies in generating 
assertions to act as oracles~\cite{Motwani19icse}.
Luckily, LLMs may be able to help.
For example, ChatAssert~\cite{Hayet25} generates test oracles using code
summaries and examples, and then uses static and dynamic analyses to repair
incorrect assertions.
%
%
%
But LLMs sometimes oversimplify specifications, and at other times fabricate them entirely~\cite{Li25Extracting}.
Thus, to effectively use LLMs to generate tests, Juan needs to review and
assess the correctness of the generated oracle assertions.
%
%

They ask an LLM to generate a test oracle for this docstring in the
form of an assert with function input \texttt{numbers} and function return \texttt{ret}. The model returns:

\begin{lstlisting}[basicstyle=\ttfamily\footnotesize]
assert not ret or any(abs(j-i) < threshold 
  for i in numbers for j in numbers if i != j)
\end{lstlisting}

This postcondition asserts that if the function returns True, then some pair of
non-equivalent elements in the input list must have an absolute difference
less than the given threshold.
This statement may seem satisfactory at first glance, but contains two
errors.
The first is an issue of completeness: it does not perform any checks
in cases where \texttt{ret} is False.
%
%
The second is an issue of correctness:
the \texttt{if i != j} check excludes all elements with equivalent values,
rather than equivalent indices.
%


Comprehending assert statements can be difficult in practice. 
%
If Juan misses the completeness error, they fail to test half of their intended functionality. 
If Juan misses the correctness error, they will end up developing a function that is entirely 
mismatched with their intended functionality. 
A natural question thus arises: if Juan prompts the LLM to also generate a comment explaining the 
test oracle, can the resulting natural language translation of the test oracle help them
catch such errors? 
Furthermore, if the LLM's comment is misaligned with the logic of the oracle (i.e., of poor quality), 
how will this misalignment affect Juan's understanding of the test? 



Unfortunately, a dearth of research exists into developer understanding of LLM-generated 
test oracles, whether LLM-generated comments can aid this understanding, and whether 
misaligned comments can hurt this understanding. 
%
%
This work aims to bridge this gap by investigating 1) how reliably developers can identify 
postcondition correctness, 2) whether LLM-generated comments can aid in this task, and 3) whether
the quality of these comments matters. We augment this investigation with a qualitative analysis 
of developer reasoning during postcondition comprehension tasks.

\section{Experimental Methodology}
\label{sec:methodology}

To understand how programming logic, natural language,
and human reasoning interact, we conducted two user studies:

\begin{enumerate}[leftmargin=*]

  \item \emph{Controlled Experiment:} A large-scale experiment
  with \quantPopSize participants. Participants assessed
  if a series of logical postconditions accurately captured the a function's behavior. We varied stimuli in a controlled manner to understand the
  impact of assertion correctness or the
  quality of an accompanying natural-language comment on human judgment.

  \item \emph{Qualitative Think-Aloud:} For a nuanced understanding
  of developer reasoning, we conducted a think-aloud with 10
  participants. Participants interacted with a series
  of postconditions and functions while verbalizing their reasoning.

\end{enumerate}

These studies combine the strengths of controlled 
analysis with insight into participant reasoning, enabling a deeper understanding of how programmers interpret
specifications.
%

\subsection{Study 1: Controlled Experiment}
\label{subsec:mainStudyDesign}

To study how assertion correctness and the presence or quality of an associated natural-language comment influence human judgement, we conducted a controlled experiment with \quantPopSize participants. Participants reviewed a series of stimuli, each consisting of a function and associated docstring, postcondition assertion, and (optionally) a natural-language comment describing the assertion. 
Participants assessed whether the postcondition accurately described the function's behavior, assuming the provided implementation was correct. 
Participants also rated the \emph{completeness} of the postcondition---that is, how fully they felt it captured the function's behavior. For both correctness and completeness judgments, participants reported confidence using a 0--5 Likert-style scale. 

Our experimental design was guided by two considerations: (1) ~balancing experimental power against participant attention, and (2)~isolating the effects of postcondition correctness and comment quality. Each participant completed 10 stimuli. We chose 10 after a small pilot study where we varied the number and found that participants began to lose attention after 10. The study was hosted on Qualtrics, took 45--60 minutes, and contained 3 attention-checks. 
Stimuli were balanced within and across participants with respect to two postcondition correctness conditions (correct and incorrect) and five comment conditions (exact, over-specified, underspecfied, incorrect, and no-comment). We discuss these conditions below.
%
%

\subsubsection{Experimental Conditions}
\label{sec:experimentalConditions}


To isolate the effects of programming logic and natural-language explanations, we systematically varied two aspects of each stimulus: (1) postcondition correctness and (2) the presence and quality of an accompanying natural-language comment.
This led to ten variants of every function (2 correctness $\times$ 5 comment conditions).

\vspace{3pt}
\noindent\emph{Correctness Conditions:} 
For each function, our corpus included two postconditions, one correct with respect to the function, and one incorrect:

\begin{enumerate}[leftmargin=*]
    \item \emph{Correct:} For all function input/output pairs that align with the requirements in the function docstring, there are no cases where the postcondition raises an error.
    \item \emph{Incorrect:} There is at least one input/output pair that aligns with requirements in the function docstring where the postcondition raises an error.
\end{enumerate}

\begin{figure*}[t]
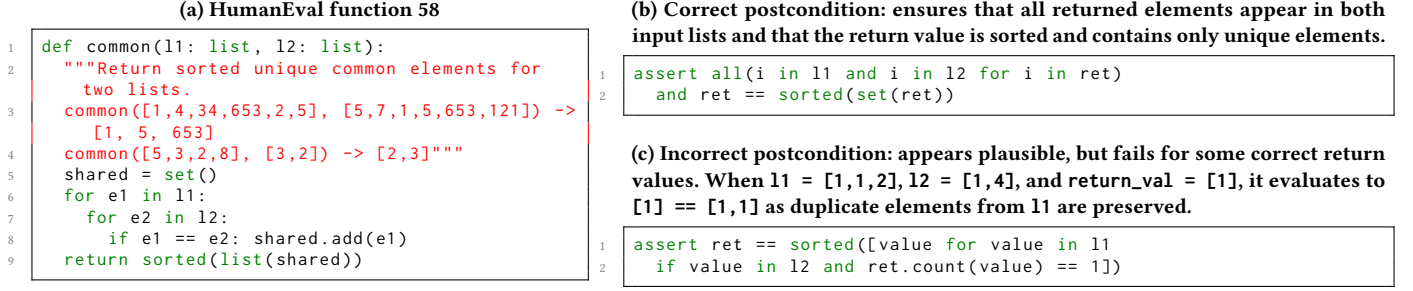

\centering

\begin{subfigure}[t]{0.401\textwidth}
\captionsetup[subfigure]{skip=0pt}
\centering
\caption{HumanEval function 58}
\label{fig:exampleFunction}
\begin{lstlisting}[basicstyle=\ttfamily\footnotesize]
def common(l1: list, l2: list):
  """Return sorted unique common elements for two lists. 
  common([1,4,34,653,2,5], [5,7,1,5,653,121]) -> [1, 5, 653]
  common([5,3,2,8], [3,2]) -> [2,3]"""
  shared = set()
  for e1 in l1:
    for e2 in l2:
      if e1 == e2: shared.add(e1)
  return sorted(list(shared))
\end{lstlisting}
\end{subfigure}
\hfill
\begin{subfigure}[t]{0.56\textwidth}
\centering

\begin{subfigure}[t]{\textwidth}
\centering
\caption{Correct postcondition: ensures that all
returned elements appear in both
input lists and that the return
value is sorted and contains only
unique elements.}
\label{subfig:correct}
\begin{lstlisting}[basicstyle=\ttfamily\footnotesize]
assert all(i in l1 and i in l2 for i in ret) 
  and ret == sorted(set(ret))
\end{lstlisting}

\end{subfigure}

\begin{subfigure}[t]{\textwidth}
\caption{Incorrect postcondition: appears
plausible, but fails for some correct
return values. When
\texttt{l1 = [1,1,2]},
\texttt{l2 = [1,4]}, and
\texttt{return\_val = [1]}, it evaluates to
\texttt{[1] == [1,1]} as
duplicate elements from
\texttt{l1} are preserved.}
\begin{lstlisting}[basicstyle=\ttfamily\footnotesize]
assert ret == sorted([value for value in l1
  if value in l2 and ret.count(value) == 1])
\end{lstlisting}

\end{subfigure}
\end{subfigure}
\caption{Example stimuli shown to participants, including the target function and associated postcondition conditions.}
\label{fig:stimulus-example}
\end{figure*}

Figure~\ref{fig:stimulus-example} has an example stimulus, including the target 
function and associated postcondition conditions. 

\vspace{3pt}
\noindent\emph{Comment Conditions:} For each postcondition--function pair, our corpus contained five stimuli:  one for each comment condition. 
%
%
In our definitions, the `failure' of an postcondition indicates that the Python
assert statement will raise an error. `passing' 
indicates that the Python assert statement will not error.
When we refer to the `failure' or `passing' of a comment, we are referring to
a direct translation of the comment into a Python assert statement.

\begin{enumerate}[leftmargin=*]
    \item \emph{Exact:} Comment is equivalent to the postcondition; For all function input/output pairs, there are no cases where the comment passes and the postcondition fails, or vice versa. We consider exact comments to be high-quality comments.

    \item \emph{Over-specified:} Comment is stricter than the postcondition; There exists a function input/output where the comment fails and the postcondition  passes, but not vice versa.

    \item \emph{Under-specified:} Comment is less strict than the postcondition; There is a function input/output where the comment passes and the postcondition fails, but not vice versa.

    \item \emph{Incorrect:} There exist function input/output pairs where the comment fails and the postcondition assertion passes, and there exist function input/output pairs where the comment passes and the postcondition assertion fails.

    \item \emph{No Comment:} No natural-language comment was included in the stimulus. The stimulus consisted entirely of the target function and candidate postcondition.
\end{enumerate}

\begin{figure*}[t]
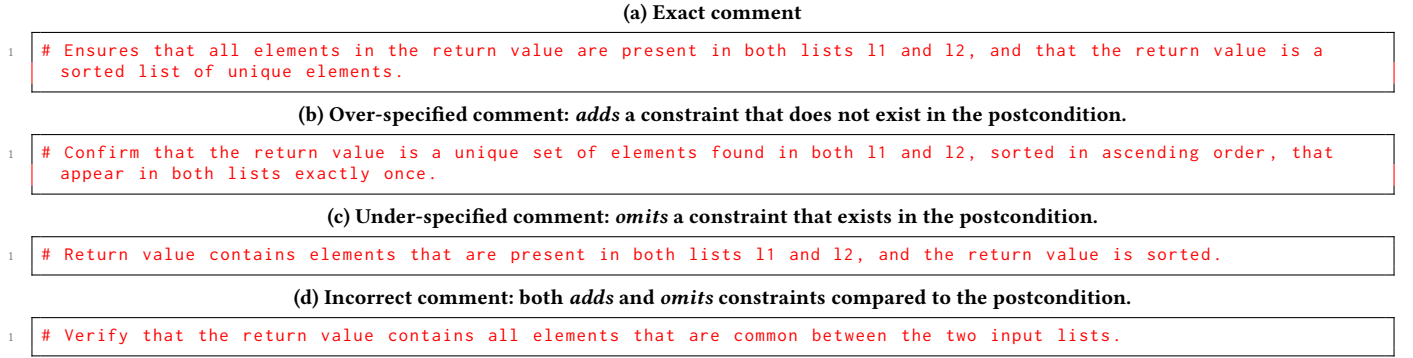

\captionsetup[subfigure]{skip=0pt}
\begin{subfigure}[t]{\textwidth}
\caption{Exact comment}
\begin{lstlisting}[basicstyle=\ttfamily\footnotesize]
# Ensures that all elements in the return value are present in both lists l1 and l2, and that the return value is a sorted list of unique elements.
\end{lstlisting}
\end{subfigure}
\begin{subfigure}[t]{\textwidth}

\caption{Over-specified comment: \emph{adds} a constraint that does not exist in the postcondition.}
\begin{lstlisting}[basicstyle=\ttfamily\footnotesize]
# Confirm that the return value is a unique set of elements found in both l1 and l2, sorted in ascending order, that appear in both lists exactly once.
\end{lstlisting}

\end{subfigure}
\begin{subfigure}[t]{\textwidth}
\caption{Under-specified comment: \emph{omits} a constraint that exists in the postcondition.}
\begin{lstlisting}[basicstyle=\ttfamily\footnotesize]
# Return value contains elements that are present in both lists l1 and l2, and the return value is sorted.
\end{lstlisting}

\end{subfigure}
\begin{subfigure}[t]{\textwidth}
\caption{Incorrect comment: both \emph{adds} and \emph{omits} constraints compared to the postcondition.}
\begin{lstlisting}[basicstyle=\ttfamily\footnotesize]
# Verify that the return value contains all elements that are common between the two input lists.
\end{lstlisting}
\end{subfigure}

\caption{Example comments corresponding to the correct postcondition in Figure~\ref{fig:stimulus-example} under each comment-quality condition.}
\label{fig:comment-examples}
\end{figure*}

Figure~\ref{fig:comment-examples} shows an example of an exact, over-specified, under-specified, and incorrect comment for the correct postcondition from Figure~\ref{fig:stimulus-example}.

\subsubsection{Stimulus Corpus Construction} 
\label{sec:stimuli}

We now describe how we constructed a corpus instantiating the experimental conditions described in Section~\ref{sec:experimentalConditions}.
%
%
Each stimulus consisted of a function, a postcondition assertion, and (optionally) a natural-language comment describing the postcondition. 

For functions, we used a subset of the HumanEval~\cite{chen2021codex} dataset developed by OpenAI for 
evaluating code generation models. We selected this dataset because it was previously 
used in prior work on LLM-generated postconditions by Endres et al.~\cite{Endres24}. In 
addition, HumanEval functions are sufficiently complex to permit non-trivial 
postconditions while remaining simple enough for inclusion in an online survey without 
overloading participants.

In their previous work, Endres et al. used GPT-3.5-turbo and GPT-4 to generate 
postcondition/comment pairs for the 164 HumanEval functions and evaluated the resulting 
postconditions for correctness and completeness~\cite{Endres24}. From this dataset, we 
filtered for functions that had at least one correct and one incorrect postcondition. We 
required that the incorrect postcondition be logically incorrect, rather than 
syntactically. We further filtered out cases where either postcondition was logically 
trivial, prohibitively long, or referenced out-of-context functions. We identified 29 
functions that fit our requirements.

All postconditions in the Endres et al. dataset included an associated LLM-generated 
comment. Some postconditions already had multiple associated comments spanning several 
quality categories because the original dataset contained repeated postconditions paired 
with different comments. When one or more comment-quality categories (exact, over-specified, under-specified, and incorrect) were missing for a 
postcondition, the first author used GPT-3.5-turbo to generate additional candidate 
comments. These candidates were manually reviewed and annotated to produce one example of 
each comment-quality condition for every postcondition.

Three authors independently reviewed all postconditions and comments for quality and correctness consistency before inclusion in our experimental corpus. At this stage, we removed three additional functions from the corpus due to implementation ambiguity in the original HumanEval dataset.

The final corpus contained 26 functions. Each function was paired with one 
correct and one incorrect postcondition, and each postcondition was paired with all five 
comment conditions, 
yielding a total of 260 experimental stimuli.

\subsection{Study 2: Qualitative Think-Aloud}
\label{subsec:followUps}

While our controlled experiment quantified programmer performance, it did not explain why participants succeeded or failed. To better understand how programmers approach and analyzed postconditions, we also conducted a series of think-aloud sessions where participants described their thought process when responding to our survey questions for 5 curated functions. 
We curated these questions by ordering our HumanEval functions by the difference between the percentage of participants who accurately identified the correct postcondition as correct, and the percentage of participants who accurately identified the incorrect postcondition as incorrect. 
We chose the 3 functions with the highest differences between these values, and the 2 functions with the lowest differences between these values. 
We counterbalanced function order, as well as whether follow-up participants were given correct or incorrect postconditions across questions, such that each participant saw 3 incorrect postconditions and 2 correct postconditions. All stimuli included exact comments. 

\section{Research Questions and Analysis Approach}
\label{sec:RQsAndAnalysis}

We organize our investigation and analysis of programming logic, natural language comments, and human comprehension
around four research questions. As described in Section~\ref{sec:methodology}, we approach our investigation from the perspective of a programmer assessing postconditions in Python.

\begin{itemize}[leftmargin=0pt,label={},itemsep=0pt,topsep=0pt,parsep=0pt,partopsep=0pt]
\item \pheading{RQ1---Postcondition Correctness:} How does the \emph{logical correctness} of an LLM-generated postcondition assertion impact programmers' ability to determine whether the assertion correctly describes the behavior of a Python function?

\item \pheading{RQ2---Natural-Language Comments:} How does the \emph{presence} and \emph{quality} of LLM-generated comments affect programmers' ability to identify whether a postcondition correctly describes the behavior of a Python function?

\item \pheading{RQ3---Assertion Characteristics:} What features of assertions (e.g., complexity or completeness) correlate with programmer performance when evaluating postconditions?

\item \pheading{RQ4---Developer Reasoning:} How do programmers reason about postconditions and accompanying natural-language comments? Are particular reasoning strategies associated with successful comprehension?
\end{itemize}

\vspace{3pt}

Although our controlled experiment was originally designed to 
investigate the effects of natural-language comments (RQ2), early 
analysis revealed a substantially larger-than-anticipated difference 
between programmers' ability to recognize correct and incorrect 
postconditions. This unexpected finding motivated additional 
analyses examining the role of postcondition correctness (RQ1), 
postcondition characteristics associated with performance 
(RQ3), and our qualitative follow-up study of the reasoning strategies underlying these differences (RQ4).

To support the original experimental goal, we pre-registered 
our hypotheses relating to RQ2 on the Open 
Science Framework~\cite{Foster17}. In particular, we hypothesized 
that natural-language comments would improve participants' ability 
to evaluate assertion correctness, but that these effects would 
depend on comment quality. All pre-registered hypotheses, study 
materials, analysis scripts, and replication artifacts are included 
in our replication package~\cite{preregistration, replication}. 
While RQ1, RQ3, and RQ4 were motivated by observations made during the early stages of analysis (and are thus more exploratory), all analyses were conducted using the same controlled experimental design and statistical methodology (see Sections~\ref{sec:methodology} and~\ref{subsec:statsMethods}).

\subsection{Quantitative and Statistical Methods} 

\label{subsec:statsMethods}

We address RQ1--RQ3 through quantitative analysis of our controlled 
experiment with 86 programmers. Across these analyses, our primary 
dependent variables are participant accuracy (whether the 
participant correctly assessed the postcondition), response time, 
and self-reported confidence.

%


Unless otherwise stated, we use mixed-effects models throughout our  analyses. We use generalized linear mixed-effects 
models for binary outcomes (e.g., participant accuracy) and linear 
mixed-effects models for continuous outcomes (e.g., response time 
and confidence). Because response times were right-skewed, we 
log-transformed them prior to analysis following established 
recommended best practice~\cite{West22,Curran18}.

%
%
%


All models include participant- and stimulus-level random effects to 
account for repeated measures,  participant ability, and variation in 
stimulus difficulty. Fixed effects were selected according to the 
research question. For example, when evaluating 
the effects of comment quality, we include postcondition correctness 
as a covariate to account for its potential influence on performance. 
%
%
%
%
We used odds ratios for generalized linear 
model effect sizes, and estimated marginal means for 
linear models. When comparing binary variables, we 
additionally report $\chi^2$ tests and Fisher's exact tests where 
appropriate.
%


We used a significance threshold of $p < 0.05$. 
We do not apply a global correction for multiple comparisons
because the primary analyses for RQ1 and the pre-registered analyses for RQ2 each centered on only three mixed-effects models: one each for accuracy, response time, and confidence. Such models jointly estimate all fixed and random effects, rather than requiring separate post hoc comparisons. 
%
%
%
Additional analyses, including the feature analyses for RQ3, were exploratory and are labeled as such in our results. 
Following recommendations for exploratory analyses~\cite{rothman1990no}, we do not apply multiple comparison corrections to reduce the likelihood of false negatives. 

We used RStudio~\cite{RStudio} for our analysis. 
R packages included \texttt{lme4}, \texttt{lmerTest}, \texttt{stats}, \texttt{car}, and \texttt{psych}. 
We include all analysis scripts and report the results of all conducted statistical tests in our replication package~\cite{replication}. 
 

\subsection{Qualitative Methods and Final Codebook} 
\label{sec:qualAnalysis}

We address RQ4 through a qualitative think-aloud study designed to 
better understand how programmers reason about logical assertions 
and accompanying natural-language comments (Section~\ref{subsec:followUps}).
Participants verbalized their reasoning while completing five assertion-evaluation tasks.
To analyze the think-aloud transcripts, we conducted a directed content analysis~\cite{hsieh2005three} using a hybrid deductive--inductive approach. 
Two authors collaboratively developed and refined a codebook capturing: (1)~how participants constructed a mental model of the function and postcondition, and (2)~the strategies participants used to determine postcondition correctness. 

Our mental-model coding scheme was informed by prior work on code comprehension~\cite{Dunsmore00,Figl25}. Specifically, we   
categorized participant--stimulus pairs as using a \emph{top-down}, \emph{bottom-up}, or \emph{as-needed} comprehension processes:

\begin{itemize}[leftmargin=0pt,label={}]
\item \emph{Top-Down}: Formed expectations from the docstring or comment, then verified them by reading the code.

\item \emph{Bottom-Up}: Inferred behavior from the code, then verified it using the docstring or comment.

\item \emph{As-Needed}: Alternated between the code and the docstring or comment as needed to resolve questions.
\end{itemize}

Correctness strategies were identified inductively as they emerged during analysis, and are one of our contributions in RQ4. To improve reliability, two authors jointly coded the transcript of one participant, and refined code definitions by resolving coding disagreements. 
The first author then applied the finalized codebook to the remaining transcripts.

\section{Participants and Recruitment}

For both studies, we recruited via computing-specific email lists and Slack channels at three large public universities. We also used snowball sampling seeded by software company contacts. 
Participants had to (1)~be at least 18, (2)~able to read and understand Python code, and (3)~correctly answer all 3 study attention checks.
Participants received a \$12 gift card.

\subsubsection*{Controlled Experiment Participants}

We collected 106 survey submissions, of which 86 met our inclusion criteria and were retained for analysis. 
We excluded participants who failed one or more attention checks. 
To remove implausibly rushed responses, we also excluded stimulus-level data completed in under 20 seconds (18/737 total stimuli); the mean time spent on a non-attention-check question was $133.5$ seconds.

Participants ranged in age between 18 and 42, with a mean age of $27.1$. $66.3\%$ of participants were professional developers. 
Most participants reported between one and ten years of programming experience, moderate or greater Python proficiency, frequent Python use, and at least some prior experience writing logical assertions and using LLMs.

%
%


\subsubsection*{Think-Aloud Participants}



To better understand the reasoning processes underlying our quantitative findings, we recruited 10 additional participants for the think-aloud study. Figure~\ref{fig:qualdems} shows participant demographics and programming experience.
Participants ranged in age from 21 to 32 years. 
Seven participants were professional programmers and three were university students. 
Nine participants had more than five years of programming experience, although only one reported significant experience writing logical assertions.

%
%
%
%


\begin{figure}
\centering
\footnotesize
\begin{tabular}{cccccccc}
\toprule
\setlength{\tabcolsep}{3pt}
\textbf{ID} & \textbf{Age} & \textbf{Prof.} & \textbf{Years} & \textbf{LLM Use} & \textbf{Writes} & \textbf{Assertion} \\
    & & \textbf{Dev.} & \textbf{Exp.} & & \textbf{Python} & \textbf{Exp.} \\
\midrule
1 & 31 & Yes & 10--15 & Sometimes       & Daily  & Some       \\
2 & 26 & No  & 10--15 & Never           & Rarely & Very little \\
3 &25 & Yes & 5--10  & Daily           & Daily  & Moderate   \\
4 &27 & Yes & 10--15 & Rarely          & Daily  & Moderate   \\
5 &32 & No  & 5--10  & Often           & Often  & Some       \\
6 &26 & No  & 5--10  & Sometimes       & Often  & Some       \\
7 &21 & Yes & 1--5   & Sometimes       & Often  & Very little \\
8 &32 & Yes & {$>$}15 & Rarely         & Often  & A lot      \\
9 &32 & Yes & 10--15 & Daily           & Rarely & Some       \\
10 & 29 & Yes & 5--10  & Daily           & Daily  & Moderate   \\
\bottomrule
\end{tabular}
\caption{Think-aloud participant overview. 
} 
\label{fig:qualdems}
\end{figure}

\section{Results}

We now present our experimental results, organized around our four research questions (see Section~\ref{sec:RQsAndAnalysis}).

\subsection{RQ1: The Impact of Assertion Correctness}

\label{rq:correctness}

Programmers were substantially better at recognizing correct postconditions than identifying incorrect ones. 
Participants correctly identified 73.9\% of correct postconditions as correct, but only 49.0\% of incorrect postconditions as incorrect. 
This difference was significant ($p < 0.001$, $OR = 2.94$), indicating that the odds of correctly recognizing a correct postcondition were nearly three times higher than those of correctly identifying an incorrect one. As a qualitative example, our stimulus example in Figure~\ref{fig:stimulus-example} is the function where we saw the largest difference in participant accuracy between the correct and incorrect postcondition ($92\%$ vs. 10\%, respectively). 
%
%

We hypothesize two potential reasons for this discrepancy. 
First, developers may have an acceptance bias toward AI-generated specifications; prior work has shown that people are likely to accept code suggestions even when they are incorrect~\cite{Eladawy24}. 
Second, it may be more cognitively demanding to identify that a logical statement is incorrect than it is to confirm that the statement is correct. 
We return to this second possibility in RQ3 (Section~\ref{rq:assertionFactors}) and RQ4 (Section~\ref{rq:developerReasoning}).

\begin{figure}[t!]
    \centering
    \includegraphics[width=\columnwidth]{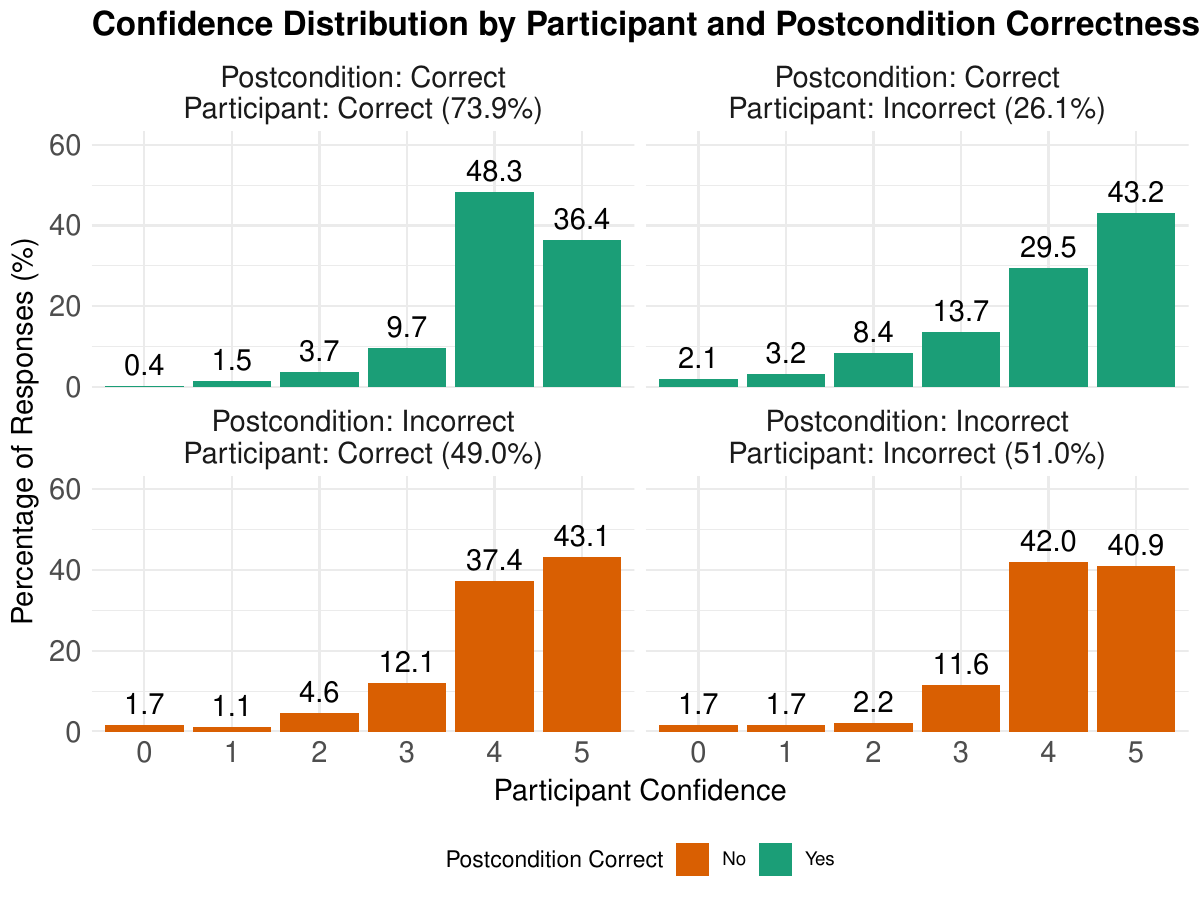}
    \caption[Participant Confidence]{Confidence distributions of participants in four cases. 
    Case 1 (top left): Participant succeeds in identifying a correct postcondition. 
    Case 2 (top right): Participant fails in identifying a correct postcondition. 
    Case 3 (bottom left): Participant succeeds in identifying an incorrect postcondition. 
    Case 4 (bottom right): Participant fails in identifying an incorrect postcondition. 
    } \label{fig:participant_confidence}
\end{figure}

Interestingly, neither postcondition correctness nor participant correctness significantly affected self-reported confidence ($p > 0.1$). Confidence was uniformly high in all cases ($3.98$--$4.14$, 0--5 Likert). 
Figure~\ref{fig:participant_confidence} summarizes confidence distributions by postcondition and participant correctness. 

Participants frequently used negative examples when they perceived a postcondition to be incorrect (18/29 cases), typically to confirm an initial judgment formed through clause comparison, logic walk-throughs, or intuition (14/18).

%
We also examined the impact of assertion correctness on response time.  
When the postcondition was correct, participants took about the same amount of time to respond whether they answered correctly or not ($103.4$ vs. $108.4$ average seconds). 
In contrast, successfully identifying an incorrect postcondition required substantially more time ($123.0$ s) than incorrectly accepting it as correct ($97.1$ s).
This difference was significant; a mixed-effects model with participant accuracy and postcondition correctness as predictors shows a significant effect of participant accuracy on response time ($p = 0.014$), driven by the interaction effect between 
participant accuracy and postcondition correctness ($p = 0.002$). 
This suggests that detecting specification errors requires effort, and incorrectly accepting plausible assertions often occurs quickly.

\summarytcolorbox{\textbf{RQ1: Summary---Impact of Assertion Correctness:} Programmers were much better at recognizing correct assertions than identifying incorrect ones. Detecting incorrect assertions also required significantly more time, yet participants remained similarly confident regardless of whether their judgments were correct.

%
%
}

\subsection{RQ2: Comment Presence and Quality}
\label{rq:comments}

We now investigate how natural-language comments impact programmer assessments of postcondition correctness. Surprisingly, we find that (1)~comments generally do not improve programer accuracy, and (2)~low-quality comments can reduce performance and increase confidence. 

We fit a mixed effects model predicting participant accuracy from comment condition while controlling for postcondition correctness. 
Contrary to our pre-registered hypothesis (see Section~\ref{sec:RQsAndAnalysis}), the presence of a comment (even if that comment was of high-quality) did not significantly improve participant accuracy compared to providing no comment at all ($p>0.1$). 
Furthermore, these effects did not differ between correct and incorrect postconditions. 
These findings suggest that simply providing an explanation is insufficient; explanation quality can matter more than explanation presence.

%


Although comments provided no overall benefit, comment quality did matter. 
When comparing between different comment qualities in the case that a comment exists, participants were significantly less accurate when comments were under-specified than when comments exactly matched the postcondition ($p = 0.037$, $OR = 0.58$). 
%
Anecdotally, participants were also slightly less accurate with under-specified comments than with no comment at all(54.3\% vs. 62.7\%), suggesting that incomplete explanations may actively interfere with reasoning rather than simply failing to help.

This decrease in performance was accompanied by increased confidence.  
Participants reported significantly higher confidence when evaluating under-specified comments than when evaluating stimuli without comments ($p = 0.005$; emmeans = 4.25/5 vs. 3.99/5). 
We found no significant relationship between participant confidence and judgment correctness. 
Thus, under-specified comments may have increased perceived understanding without improving actual comprehension.

%

Comment condition also affected response time ($p = 0.022$). 
Participants took the longest time to respond when looking at exact comments, and the shortest time to respond when there was no comment at all (\emph{emmeans} $=$ $94.3$, $114.5$, $105.3$, $109.6$, $111.8$ for no comment, exact comment, incorrect comment, over-specified comment, 
and under-specified comment, respectively). 
Participants may have spent additional time comparing the comment with the postcondition, although this additional effort did not improve judgment accuracy.

\summarytcolorbox{\textbf{RQ2: Summary---Impact of Comments:} Contrary to our pre-registered hypotheses, natural-language comments did not improve programmers' ability to evaluate postconditions.
%
%
Furthermore, under-specified comments reduced accuracy relative to exact comments, while increasing developer confidence compared to no comment at all.
}

\subsection{RQ3: Assertion Characteristics and Performance}

\label{rq:assertionFactors}

Because the large asymmetry observed in RQ1 was unexpected, we conducted several exploratory analyses to better understand what characteristics of postcondition assertions were associated with participant performance.

\vspace{2pt}
\noindent\textbf{Completeness:} 
For correct postconditions, we compared participant completeness ratings against the independently computed mutation-based bug-completeness scores from Endres et al.~\cite{Endres24}.
Participants' completeness ratings showed a significant positive correlation with bug-completeness (Spearman's $\rho = 0.18$, $p < 0.001$). 
Participants also rated correct postconditions as significantly more complete than incorrect ones (Wilcoxon rank-sum test, $p < 0.001$; \emph{means} = $=$ $2.66$ vs. $2.35$). 
Together, these findings suggest that participants could distinguish stronger specifications from weaker ones even when they failed to identify logical errors. 
 
%
%
%
%


\vspace{2pt}
\noindent\textbf{Complexity:} Correct postconditions exhibited significantly lower cyclomatic complexity than incorrect postconditions ($p < 0.001$; \emph{emmeans} = 4.28 vs. 4.85), a pattern that we find is also true across the larger postcondition corpus from prior work~\cite{Endres24} from which we filter for our experimental stimuli. 
However, mediation analysis found that cyclomatic complexity did not explain the effect of postcondition correctness on participant accuracy ($p = 0.50$), 
and neither Halstead difficulty nor cognitive complexity differed significantly between correct and incorrect postconditions ($p > 0.1$). 
Thus, although incorrect assertions tend to be somewhat more complex, complexity alone does not explain developers' difficulty detecting them.

%

%
%

\vspace{2pt}
\noindent\textbf{Assertion Construction Patterns:} Using the postcondition taxonomy proposed by Endres et al.~\cite{Endres24}, we explored whether particular assertion construction patterns were associated with participant performance.
%
%

%

Assertions containing type checks ($\chi^2 = 7.99$, $p = 0.005$, $OR = 2.10$) or arithmetic equalities ($\chi^2 = 9.64$, $p = 0.002$, $OR = 2.07$) were more likely to be judged correctly, whereas implications ($\chi^2 = 3.84$, $p = 0.05$, $OR = 0.69$) and element-property assertions ($\chi^2 = 3.26$, $p = 0.07$, $OR = 0.66$) were associated with decreased performance. 
We return to these observations in RQ4 (Section~\ref{rq:developerReasoning}), where our qualitative analysis suggests these construction patterns differ in the extent to which they support direct clause comparison.

\subsubsection{Demographics} 
Finally, we found no evidence that developer experience, Python expertise, assertion-writing experience, or LLM usage predicted participant accuracy.
This may indicate that programmers of all experience levels could benefit from assistance with parsing postconditions.

\summarytcolorbox{\textbf{RQ3: Summary---Assertion Construction Patterns:} Participants could recognize differences in completeness between postconditions.
Although incorrect assertions tended to be slightly more complex, complexity did not explain the performance difference observed in RQ1. Some logical structures are easier for developers to evaluate than others.
}

\subsection{RQ4: Understanding Developer Reasoning}

\label{rq:developerReasoning}

To provide insight into developer reasoning, we conduct a qualitative think-aloud, where we investigate (1) how participants understand functions and associated postcondition assertions, and (2) how do participants reason about postcondition correctness. We collected think-aloud transcripts for 50 function--postcondition pairs (10 participants, 5 stimuli each). Figure~\ref{fig:function_performance} describes these stimuli, which were selected to cover a range of reasoning difficulties (as determined by participant performance in our large quantitative evaluation).

\begin{figure}[t]
\centering
\small
\begin{tabular}{lccc}
\toprule
& \multicolumn{2}{c}{\textbf{\% Identified}} & \\
\cmidrule(lr){2-3}
\textbf{Function} &
\textbf{Incorrect} &
\textbf{Correct} &
\textbf{Diff} \\
\midrule
common              & 10.0  & 92.3 & 82.3 \\
match\_parens       & 82.4  & 18.2 & 64.2 \\
has\_close\_elements& 17.6  & 76.9 & 59.3 \\
any\_int            & 75.0  & 82.4 & 7.4 \\
choose\_num         & 0.6   & 0.611 & 1.1 \\
\bottomrule
\end{tabular}
\caption{Functions curated for our qualitative follow-up study.
These functions make up the top three and the bottom two of our stimuli,
ordered by the difference between the
percentage of participants who accurately identified the correct
postcondition as correct and those who
accurately identified the incorrect postcondition as incorrect.}
\label{fig:function_performance}
\end{figure}

\vspace{2pt}
\noindent\textbf{Comprehension: Building a mental model:} As described in Section~\ref{sec:qualAnalysis}, we code each interaction for its predominant \emph{comprehension process} (top-down, bottom-up, or as-needed).

For the majority of responses (31/50), participants used a top-down approach, first reading the English language docstring or comment, and then confirming their understanding by parsing the code.
In 7 responses, participants went bottom-up, first reading the code and then confirming with comments. In 10 responses, participants showcased an as-needed approach, bouncing between the documentation and the code. 
In 2 responses, the participants did not provide enough spoken information for us to make a determination.

\vspace{2pt}
\noindent\textbf{Reasoning Strategies:} We identified five strategies used by participants to determine postcondition correctness: 

\begin{enumerate}[leftmargin=*]
\item \emph{Clause Comparison}: The participant compared logical clauses between the function and the postcondition, and determined whether they matched. \qualquote{Now, looking at this assertion, it's entirely missing the integer criterion.}{4}

\item \emph{Logic Walk-through}: The participant traced the flow of data through the postcondition to determine whether it matched an expected data flow with respect to the function. 
\qualquote{This is just saying, return whatever the maximum value is in that valid range if X is less than or equal to Y \ldots 
otherwise negative 1. It seems correct.}{3}

\item \emph{Positive Examples}: The participant walked through a specific example of an input-output pair for the function that passed the postcondition. 
\qualquote{Empty list, that would be fine. [The postcondition] would be vacuously satisfied if the return value is just one thing.}{4}

\item \emph{Negative Examples}: The participant walked through a specific example of an input-output pair for the function that failed the postcondition. 
\qualquote{If X and Y were 3 and 3, then the return value [of the function] would be negative 1, but ... this assertion would fail.}{10}

\item \emph{Intuition}: The participant did not systematically determine postcondition correctness, but rather intuited that the postcondition seemed correct or incorrect. 
\qualquote{I'm not totally certain, maybe I'm missing some subtle detail, but it sounds correct. It seems like it's capturing the correct behavior... 
I can't think of how I'm not right.}{3}

\end{enumerate}

\begin{figure}[t!]
    \centering
    \includegraphics[width=0.9\columnwidth]{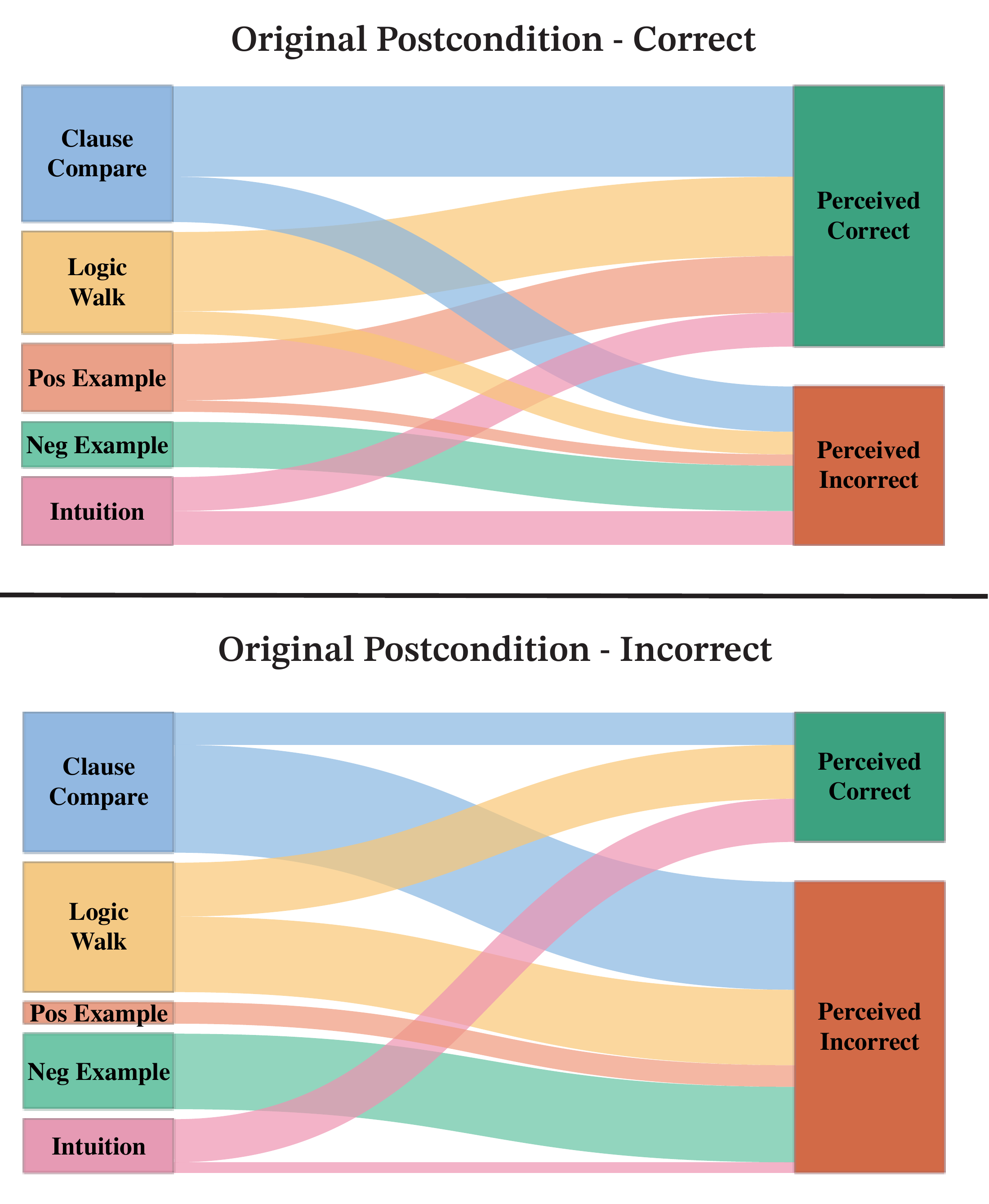}
    \caption[Think-Aloud Strategies]{Two Sankey diagrams showing the 
    portion of responses from the think-aloud survey where participants 
    employed each strategy when they perceived a postcondition to be 
    incorrect, and when they perceived a postcondition to be correct. 
    The diagrams are separated by the ground-truth correctness of the 
    original postconditions.}
    \label{fig:Perception_Sankeys}
\end{figure}

%
Participants often used multiple strategies to assess the same stimulus. Figure~\ref{fig:Perception_Sankeys} displays two Sankey diagrams summarizing strategy use across our five think-aloud functions (Figure~\ref{fig:function_performance}), split by when the participant perceived a postcondition to be correct and when they perceived it to be incorrect.
The top diagram summarizes responses for ground-truth correct postconditions, while the bottom diagram summarizes responses for ground-truth incorrect postconditions.

We found that in most cases where a postcondition was perceived to be incorrect (18/29), participants used negative examples. This was often to confirm a perception formed by clause comparisons, logic walk-through, or intuition (14/18).
For example, one participant viewing the \texttt{choose\_num} function intuited that \qualquoteNoPID{the assert doesn't look right}, then followed a negative example to \qualquote{think through the whole thing}{9}. 
Occasionally, participants would attempt to confirm or break their correctness perception by trying to find a counterexample. 
For instance, one participant stated that the correct postcondition for function \texttt{match\_parens} \qualquote{seems right}{10}, then went on to look for counterexamples with \qualquote{let me think if I can break this}{10}.
%

Easier postconditions tended to afford clause comparison. 
In four of the five postconditions with at least 75\% accuracy in the original study (Figure~\ref{fig:function_performance}), every think-aloud participant used clause comparison at least once.
The \texttt{any\_int} function had the highest average response accuracy across ground-truth correct and incorrect postconditions.
For this function, both postconditions afforded direct clause comparison.
For example, when viewing the incorrect postcondition, P1 noted that \qualquoteNoPID{we are missing the requirement that they are all integers}.
When viewing the correct postcondition, P6 noted that \qualquoteNoPID{it seems like this assertion checks all three conditions here}.

We observed no clear relationship between the number of strategies used and response correctness, nor with stimulus. 
However, we did observe individual differences in reasoning styles: some consistently used only one or two strategies, whereas others regularly applied three or more.
For instance, P2 used only 1 strategy in 4 out of 5 of their responses, while participant 8 used 4 or more strategies in 4 out of 5 of their responses.
Additionally, certain participants seemed to prefer certain strategies.
For instance, P3 always began with a logic walk-through,  P6 leaned on clause comparisons, and P8 consistently used examples.

%
Only 12/50 responses involved the use of intuition, with just 3 responses consisting entirely of intuition.
However, when intuition was used, participants were wrong more often than they were right, with 7/12 incorrect responses.
These made up the majority of the 11 incorrect responses.

\summarytcolorbox{\textbf{RQ4: Summary---Developer Reasoning:} Participants primarily used top-down comprehension, relying on clause comparison, logic walk-throughs, positive and negative examples, and intuition to assess function--postcondition alignment.
Participants more accurately classified postconditions that afforded clause comparison than those that did not.
Negative examples were primarily used to confirm suspected incorrectness after other reasoning strategies.}

\section{Discussion and Implications}
\label{sec:discussion}

Our findings have several implications for tool design.

\pheading{Developers tend to trust ML output even when it is wrong.} 
Humans are significantly worse at recognizing when postconditions are incorrect than when they are correct; that is, they are far more likely to return false positives than false negatives. 
Furthermore, they trust with high confidence in both cases (see Figure~\ref{fig:participant_confidence}). 
This aligns with previous findings regarding code repair suggestions~\cite{Eladawy24}, and indicates that development assistance tools should facilitate developer comprehension of suggested test oracles.

\pheading{Comments are insufficient to explain postconditions to developers.} 
Comments are a standard way to increase code comprehension~\cite{Huang20} and facilitate LLM-based development~\cite{Li26}. 
However, our findings in Section~\ref{rq:comments} indicate that comprehension of postconditions specifically may not increase significantly with comments, and that comments that are insufficiently specified may even harm comprehension. 
Future work should investigate alternative approaches to aligning human mental models to logical postcondition statements.

\pheading{Developers have an intuitive understanding of completeness.}
Humans can correctly gauge relative postcondition completeness, and can intuit that correct postconditions are more complete than incorrect postconditions (Section~\ref{rq:assertionFactors}). 
This could indicate that intuited bug-completeness can potentially help ground mental models of postcondition correctness, and should be investigated in future work.

\pheading{Incorrect postconditions tend towards higher cyclomatic complexity than correct postconditions.} 
While we did not find that this mediated our observed difference between participant tendency to identify correct and incorrect postconditions (Section~\ref{rq:assertionFactors}), this is a noteworthy difference observed not only in our filtered dataset but over the entire 
unfiltered dataset from previous work~\cite{Endres24}. 
Future work should investigate the use of cyclomatic complexity to flag potentially incorrect postconditions or inform recommendations to developers.

\section{Related Work}

%
While recent work has made substantial progress in generating specifications, assertions, and natural-language explanations, comparatively little is known about if developers can reliably evaluate these generated reliability artifacts. 
Our work helps fill this gap by systematically investigating the human ability to assess machine-generated assertions and the effect of accompanying natural-language explanations on that ability.

\subsection{Natural Language Support for AI-Generated Software}

Collaboration between developers and LLMs is now standard in development workflows~\cite{StackOverflow2025Survey}. 
%
Although LLMs perform well on many benchmarks, they remain susceptible to vulnerabilities~\cite{Gulmez26}, hallucinated APIs~\cite{Jones24, Sarkar24}, and misinterpretation of natural-language instructions~\cite{Tian25}. 
Generated code may be fundamentally misaligned with developer intent, and that misalignment can be hard to
detect
~\cite{Vaithilingam22}. 
Humans also cede agency to artificial intelligence in a way they do not cede it to humans~\cite{Candrian22}, and readily accept incorrect code~\cite{Eladawy24}.

Natural-language comments are the standard mechanism for improving code comprehension~\cite{Huang20} and can increase developer adoption of LLM-generated code~\cite{Li26}. 
%
%
Early work generated comments using static-analysis heuristics and templates~\cite{Haiduc10, Eddy13, Sridhara10, McBurney14, Abid15}. 
More recent work formulates comment generation as machine translation from code to natural language~\cite{Iyer16, Xing18, Sharma22, Huang20, Feng20, Ahmad21, Wang21, Phan21, Allamanis16, Khan23, Wan18, Gao23, Liu21, Parvez21}. 
These approaches are typically evaluated using automatic metrics such as BLEU~\cite{Papineni02}, or ROUGE~\cite{Lin04}, 
although such metrics do not necessarily correlate with developer understand, and humans may prefer human-written summaries to machine-written ones~\cite{Stapleton20}. 


As a result, several works evaluate generated comments through human studies
of how the comments impact comprehension.  
Prior work has measured comprehension using short-answer questions~\cite{Stapleton20, Grandel26}, multiple-choice questions~\cite{Wu26}, cloze tests~\cite{Borstler16}, and code writing/debugging tasks~\cite{Grandel26}. 
ComCat shows that expertise-guided context improves both generated comments and developer understanding~\cite{Grandel26}.  
%
%
B"{o}rstler et al. find that comments conveying information beyond direct code translation, such as purpose, improve readability but not necessarily comprehension~\cite{Borstler16}. 
Steidl et al. propose and validate a model that evaluates comment quality based on coherence, usefulness, completeness, and consistency~\cite{Steidl13}.
Finally, Wu et al. show that annotations containing formally verified assertions can themselves improve program understanding ~\cite{Wu26}. 

In our work, we are interested in programmers' ability to correctly assess logical assertions. Rather than evaluating generated comments through programmer comprehension tasks, we investigate how comment quality affects programmers' judgement of generated assertions.
Because assertions have precise formal semantics, we can systematically characterize comment--assertion alignment as exact, over-specified, under-specified, or incorrect, enabling controlled study of how different forms of comment quality influence programmer judgment.

\subsection{Aligning Natural Language Intent with Code}

A common approach to ensuring alignment between developer intent and generated code is to formalize human-written natural-language specifications. 
Lahiri argues that future AI-assisted software development should increasingly rely on behavioral artifacts such as tests, assertions, and formal specifications to capture developer intent~\cite{lahiri2026intent}.
Helping realize this vision, Prasad et al. propose a cognitively grounded framework that helps developers verify AI-generated code by reasoning over machine-generated formal artifacts~\cite{prasad_et_al:LIPIcs.ECOOP.2026.22}. 
Together, these works motivate specification-centric, human-in-the-loop development workflows.
However, comparatively little is known about how effectively developers can evaluate the generated specifications they are asked to review.

A large body of work has focused on generating 
formal specifications from natural-language documentation. 
ALICS~\cite{Pandita12} and Doc2Spec~\cite{Zhong09} infer formal language specifications from API documentation.
Toradocu~\cite{Goffi16}, JDoctor~\cite{Blasi18}, and C2S~\cite{Zhai20} use comments to generate oracles as precondition or postcondition assertions. 
Swami generates executable tests from structured natural-language specifications~\cite{Motwani19icse}. 
%
%
More recent work leverages LLMs, using context gathered from existing code~\cite{Molinelli25}, documentation~\cite{Hayet25, Li25Extracting}, or GitHub issues~\cite{Ruan25} to generate specifications. 

In practice, natural-language documentation or developer intent can be misaligned with implementation. Thus, other approaches attempt to catch or correct such misalignment. 
iComment extracts implicit program rules from comments and analyzes them for implementation inconsistencies~\cite{Tan07}. 
@TComment generates tests for Javadoc properties to find discrepancies between comments and code~\cite{Tan12}. 
JavadocMiner uses heuristic-based metrics to assess inline documentation quality and code-alignment~\cite{Khamis10}. 
Zhou et al. extract logical directives from API documents and use a constraint solver to detect defects~\cite{Zhou17}. 
Specine repairs misalignment between LLM-generated specifications and intent~\cite{Tian25}. 
Endres et al. propose empirically and qualitatively validated metrics to measure alignment between natural-language generated postconditions and intent~\cite{Endres24}.
Finally, TiCoder partially formalizes intent by interactively guiding developers via test generation~\cite{Fakhoury24}.

Accross this literature, evaluation of alignment between natural language and formal specifications is largely automatic. 
While recent work has proposed systems that support developers in reviewing generated specifications and formal artifacts~\cite{prasad_et_al:LIPIcs.ECOOP.2026.22}, there is limited empirical understanding of how accurately programmers can assess specification correctness or how they reason about these artifacts. 
In our study, we provide participants with both human-written intent in the form of function docstrings, and machine-generated assertion statements. 
By measuring both participant accuracy at identifying assertion correctness and perceptions of completeness, we provide empirical evidence about the human verification capability underlying  specification-driven development.



\section{Limitations and Threats to Validity}
\pheading{The population.} Our participants were recruited exclusively from a population of university students and software company employees in the United States. 
As such, we were only able to capture participant behavior in this demographic, and it is possible that other demographics will exhibit different understanding patterns of LLM-generated comments or assertions. 

\pheading{The generation model.} We used GPT-3.5-turbo to generate assertion comments, and the postcondition assertions we used were also generated by this model during previous work~\cite{Endres24}. 
More recent models have since been released, with updated code generation capabilities.
However, as our work targets human understanding rather than model capability, we posit that our results will generalize to more recent models.

\pheading{The survey dataset.} All of our survey questions, including LLM-generated comments and postconditions, were built with the HumanEval dataset~\cite{chen2021codex}. 
We chose this dataset as it was written specifically to act as a benchmark for LLM-driven code generation. 
However, this dataset includes only single-function Python code blocks with functionality that can be mostly summarized in a single assertion statement. 
We cannot be certain that our results generalize to multi-function code or to multiple assertions, though we posit that more complex code will exacerbate the concerns we have surfaced.

\section{Conclusion}

We perform a systematic evaluation of humans' ability to comprehend LLM-generated logical Python postcondition statements, and the effect of associated natural-language comments of varying quality on this comprehension.
We further perform an exploratory analysis of postcondition characteristics that affect human comprehension.
Our work contributes to the under-studied area of human understanding of LLM-generated logical specifications. 

\section*{Data Availability}

%
Our stimuli, pre-registration, data, analysis scripts, and results are available at a repository on the Open Science Framework (\url{osf.io})~\cite{replication}.


\begin{acks}
This work is supported by the National Science Foundation under grant no. CCF-2210243.
\end{acks}

\bibliographystyle{ACM-Reference-Format}
\bibliography{softeng}

@misc{preregistration, 
title={The impact of comments on the correct understanding of logical code statements}, 
url={osf.io/5f73u}, 
publisher={OSF}, 
author={Zhanna Kaufman and Yuriy Brun and Adithya Murali and Madeline Endres}, 
year={2025}, 
month={Dec}
}

@misc{replication,
  title        = {Replication Package for The impact of comments on the correct understanding of logical code statements},
  year         = 2026,
  authors      = {Zhanna Kaufman and Yuriy Brun and Adithya Murali and Madeline Endres},
  howpublished = {\url{https://osf.io/q6d2m/overview?view_only=442a1b8d82554de0baa4b4083987a6b6}}
}

@misc{Ahmad21,
      title={Unified Pre-training for Program Understanding and Generation}, 
      author={Wasi Uddin Ahmad and Saikat Chakraborty and Baishakhi Ray and Kai-Wei Chang},
      year={2021},
      eprint={2103.06333},
      archivePrefix={arXiv},
      primaryClass={cs.CL},
      url={https://arxiv.org/abs/2103.06333}, 
}

@INPROCEEDINGS{Cihan25,

  author = {Cihan, Umut and Haratian, Vahid and {\.I}\c{c}\"{o}z, Arda and G\"{u}l, Mert Kaan and Devran, \"{O}mercan and Bayendur, Emircan Furkan and U\c{c}ar, Baykal Mehmet and T\"{u}z\"{u}n, Eray},

  booktitle={International Conference on Software Engineering: Software Engineering in Practice (ICSE-SEIP)}, 

  title={Automated Code Review in Practice}, 

  year={2025},

  volume={},

  number={},

  pages={425-436},

  doi={10.1109/ICSE-SEIP66354.2025.00043}}

@inproceedings{Zhong24,
author = {Zhong, Li and Wang, Zilong},
title = {Can {LLM} replace stack overflow? a study on robustness and reliability of large language model code generation},
year = {2024},
isbn = {978-1-57735-887-9},
publisher = {AAAI Press},
url = {https://doi.org/10.1609/aaai.v38i19.30185},
doi = {10.1609/aaai.v38i19.30185},
articleno = {2437},
numpages = {9},
series = {AAAI'24/IAAI'24/EAAI'24}
}

@inproceedings{Misra20,
author = {Misra, Vishal and Reddy, Jakku Sai Krupa and Chimalakonda, Sridhar},
title = {Is there a correlation between code comments and issues? an exploratory study},
year = {2020},
isbn = {9781450368667},
publisher = {Association for Computing Machinery},
address = {New York, NY, USA},
url = {https://doi.org/10.1145/3341105.3374009},
doi = {10.1145/3341105.3374009},
booktitle = {Proceedings of the 35th Annual ACM Symposium on Applied Computing},
pages = {110–117},
numpages = {8},
location = {Brno, Czech Republic},
series = {SAC '20}
}

@misc{Kang24,
      title={Identifying Inaccurate Descriptions in {LLM}-generated Code Comments via Test Execution}, 
      author={Sungmin Kang and Louis Milliken and Shin Yoo},
      year={2024},
      eprint={2406.14836},
      archivePrefix={arXiv},
      primaryClass={cs.SE},
      url={https://arxiv.org/abs/2406.14836}, 
}

@article{Fakhoury24,
   title={LLM-Based Test-Driven Interactive Code Generation: User Study and Empirical Evaluation},
   volume={50},
   ISSN={2326-3881},
   url={http://dx.doi.org/10.1109/TSE.2024.3428972},
   DOI={10.1109/tse.2024.3428972},
   number={9},
   journal={IEEE Transactions on Software Engineering},
   publisher={Institute of Electrical and Electronics Engineers (IEEE)},
   author={Fakhoury, Sarah and Naik, Aaditya and Sakkas, Georgios and Chakraborty, Saikat and Lahiri, Shuvendu K.},
   year={2024},
   month=sep, pages={2254–2268} }

@inproceedings{Zi25,
author = {Zi, Yangtian and Li, Luisa and Guha, Arjun and Anderson, Carolyn and Feldman, Molly Q},
title = {``I Would Have Written My Code Differently': Beginners Struggle to Understand {LLM}-Generated Code},
year = {2025},
isbn = {9798400712760},
publisher = {Association for Computing Machinery},
address = {New York, NY, USA},
url = {https://doi.org/10.1145/3696630.3731663},
doi = {10.1145/3696630.3731663},
booktitle = {International Conference on the Foundations of Software Engineering},
pages = {1479–1488},
numpages = {10},
location = {Clarion Hotel Trondheim, Trondheim, Norway},
series = {FSE Companion '25}
}

@inproceedings{Vaithilingam22,
author = {Vaithilingam, Priyan and Zhang, Tianyi and Glassman, Elena L.},
title = {Expectation vs. Experience: Evaluating the Usability of Code Generation Tools Powered by Large Language Models},
year = {2022},
isbn = {9781450391566},
publisher = {Association for Computing Machinery},
address = {New York, NY, USA},
url = {https://doi.org/10.1145/3491101.3519665},
doi = {10.1145/3491101.3519665},
booktitle = {Extended Abstracts of the 2022 CHI Conference on Human Factors in Computing Systems},
articleno = {332},
numpages = {7},
location = {New Orleans, LA, USA},
series = {CHI EA '22}
}

@ARTICLE{Xia18,

  author={Xia, Xin and Bao, Lingfeng and Lo, David and Xing, Zhenchang and Hassan, Ahmed E. and Li, Shanping},

  journal={IEEE Transactions on Software Engineering}, 

  title={Measuring Program Comprehension: A Large-Scale Field Study with Professionals}, 

  year={2018},

  volume={44},

  number={10},

  pages={951-976},

  doi={10.1109/TSE.2017.2734091}}

@inproceedings{Hu22,
author = {Hu, Xing and Xia, Xin and Lo, David and Wan, Zhiyuan and Chen, Qiuyuan and Zimmermann, Thomas},
title = {Practitioners' expectations on automated code comment generation},
year = {2022},
isbn = {9781450392211},
publisher = {Association for Computing Machinery},
address = {New York, NY, USA},
url = {https://doi.org/10.1145/3510003.3510152},
doi = {10.1145/3510003.3510152},
booktitle = {Proceedings of the 44th International Conference on Software Engineering},
pages = {1693–1705},
numpages = {13},
location = {Pittsburgh, Pennsylvania},
series = {ICSE '22}
}

@article{chen2021codex,
  title={Evaluating Large Language Models Trained on Code},
  author={Mark Chen and Jerry Tworek and Heewoo Jun and Qiming Yuan and Henrique Ponde de Oliveira Pinto and Jared Kaplan and Harri Edwards and Yuri Burda and Nicholas Joseph and Greg Brockman and Alex Ray and Raul Puri and Gretchen Krueger and Michael Petrov and Heidy Khlaaf and Girish Sastry and Pamela Mishkin and Brooke Chan and Scott Gray and Nick Ryder and Mikhail Pavlov and Alethea Power and Lukasz Kaiser and Mohammad Bavarian and Clemens Winter and Philippe Tillet and Felipe Petroski Such and Dave Cummings and Matthias Plappert and Fotios Chantzis and Elizabeth Barnes and Ariel Herbert-Voss and William Hebgen Guss and Alex Nichol and Alex Paino and Nikolas Tezak and Jie Tang and Igor Babuschkin and Suchir Balaji and Shantanu Jain and William Saunders and Christopher Hesse and Andrew N. Carr and Jan Leike and Josh Achiam and Vedant Misra and Evan Morikawa and Alec Radford and Matthew Knight and Miles Brundage and Mira Murati and Katie Mayer and Peter Welinder and Bob McGrew and Dario Amodei and Sam McCandlish and Ilya Sutskever and Wojciech Zaremba},
  year={2021},
  eprint={2107.03374},
  archivePrefix={arXiv},
  primaryClass={cs.LG}
}

@inproceedings{Papineni02,
author = {Papineni, Kishore and Roukos, Salim and Ward, Todd and Zhu, Wei-Jing},
title = {BLEU: a method for automatic evaluation of machine translation},
year = {2002},
publisher = {Association for Computational Linguistics},
address = {USA},
url = {https://doi.org/10.3115/1073083.1073135},
doi = {10.3115/1073083.1073135},
booktitle = {Proceedings of the 40th Annual Meeting on Association for Computational Linguistics},
pages = {311–318},
numpages = {8},
location = {Philadelphia, Pennsylvania},
series = {ACL '02}
}

@article{Endres24,
  author = {Madeline Endres and Sarah Fakhoury and Saikat Chakraborty and Shuvendu K. Lahiri},
  title = {Can Large Language Models Transform Natural Language Intent into Formal Method Postconditions?},
  year = {2024},
  address = {Galinhas, Brazil},
  volume = {1},
  number = {FSE},
  doi = {10.1145/3660791},
  journal = {PACMSE},
  pages = {84:1--84:24},
}

@article{Foster17,
author = {Foster, Erin and Deardorff, Ariel},
year = {2017},
month = {04},
pages = {},
title = {Open Science Framework (OSF)},
volume = {105},
journal = {Journal of the Medical Library Association},
doi = {10.5195/JMLA.2017.88}
}

@article{Dunsmore00,
title = {The role of comprehension in software inspection},
journal = {Journal of Systems and Software},
volume = {52},
number = {2},
pages = {121-129},
year = {2000},
issn = {0164-1212},
doi = {https://doi.org/10.1016/S0164-1212(99)00138-7},
url = {https://www.sciencedirect.com/science/article/pii/S0164121299001387},
author = {A Dunsmore and M Roper and M Wood}
}

@article{Figl25,
author = {Figl, Kathrin and Kirchner, Maria and Baltes, Sebastian and Felderer, Michael},
title = {The Influence of Code Comments on the Perceived Helpfulness of Stack Overflow Posts},
year = {2025},
issue_date = {Oct 2025},
publisher = {Kluwer Academic Publishers},
address = {USA},
volume = {30},
number = {6},
issn = {1382-3256},
url = {https://doi.org/10.1007/s10664-025-10727-w},
doi = {10.1007/s10664-025-10727-w}
}

@article{Gulmez26,
	author = {G{\"u}lmez, Burak},
	date = {2026/04/10},
	doi = {10.1007/s10489-026-07230-0},
	id = {G{\"u}lmez2026},
	isbn = {1573-7497},
	journal = {Applied Intelligence},
	number = {6},
	pages = {200},
	title = {Code generation with large language models: a survey from neural program synthesis to autonomous software development},
	url = {https://doi.org/10.1007/s10489-026-07230-0},
	volume = {56},
	year = {2026}}

@inproceedings{Xing18,
author = {Hu, Xing and Li, Ge and Xia, Xin and Lo, David and Jin, Zhi},
title = {Deep code comment generation},
year = {2018},
isbn = {9781450357142},
publisher = {Association for Computing Machinery},
address = {New York, NY, USA},
url = {https://doi.org/10.1145/3196321.3196334},
doi = {10.1145/3196321.3196334},
booktitle = {Proceedings of the 26th Conference on Program Comprehension},
pages = {200–210},
numpages = {11},
location = {Gothenburg, Sweden},
series = {ICPC '18}
}

@inproceedings{Zhai20,
author = {Zhai, Juan and Shi, Yu and Pan, Minxue and Zhou, Guian and Liu, Yongxiang and Fang, Chunrong and Ma, Shiqing and Tan, Lin and Zhang, Xiangyu},
title = {C2S: translating natural language comments to formal program specifications},
year = {2020},
isbn = {9781450370431},
publisher = {Association for Computing Machinery},
address = {New York, NY, USA},
url = {https://doi.org/10.1145/3368089.3409716},
doi = {10.1145/3368089.3409716},
pages = {25–37},
numpages = {13},
location = {Virtual Event, USA},
series = {ESEC/FSE 2020}
}

@inproceedings{Sharma22,
author = {Sharma, Rishab and Chen, Fuxiang and Fard, Fatemeh},
title = {LAMNER: code comment generation using character language model and named entity recognition},
year = {2022},
isbn = {9781450392983},
publisher = {ACM},
address = {New York, NY, USA},
url = {https://doi.org/10.1145/3524610.3527924},
doi = {10.1145/3524610.3527924},
booktitle = {International Conference on Program Comprehension},
pages = {48–59},
numpages = {12},
location = {Virtual Event},
series = {ICPC '22}
}

@article{Huang20,
title = {Towards automatically generating block comments for code snippets},
journal = {Information and Software Technology},
volume = {127},
pages = {106373},
year = {2020},
issn = {0950-5849},
doi = {https://doi.org/10.1016/j.infsof.2020.106373},
url = {https://www.sciencedirect.com/science/article/pii/S0950584920301427},
author = {Yuan Huang and Shaohao Huang and Huanchao Chen and Xiangping Chen and Zibin Zheng and Xiapu Luo and Nan Jia and Xinyu Hu and Xiaocong Zhou},
}

@misc{Tian25,
      title={Aligning Requirement for Large Language Model's Code Generation}, 
      author={Zhao Tian and Junjie Chen},
      year={2025},
      eprint={2509.01313},
      archivePrefix={arXiv},
      primaryClass={cs.SE},
      url={https://arxiv.org/abs/2509.01313}, 
}

@inproceedings{evalplus,
  title = {Is Your Code Generated by Chat{GPT} Really Correct? Rigorous Evaluation of Large Language Models for Code Generation},
  author = {Liu, Jiawei and Xia, Chunqiu Steven and Wang, Yuyao and Zhang, Lingming},
  booktitle = {Thirty-seventh Conference on Neural Information Processing Systems},
  year = {2023},
  url = {https://openreview.net/forum?id=1qvx610Cu7},
}

@Manual{RStudio,
title = {RStudio: Integrated Development Environment for R},
author = {{Posit team}},
organization = {Posit Software, PBC},
address = {Boston, MA},
year = {2025},
url = {http://www.posit.co/},
}

@inproceedings{Jones24,
title={Teaching Language Models to Hallucinate Less with Synthetic Tasks},
author={Erik Jones and Hamid Palangi and Clarisse Sim{\~o}es Ribeiro and Varun Chandrasekaran and Subhabrata Mukherjee and Arindam Mitra and Ahmed Hassan Awadallah and Ece Kamar},
booktitle={The Twelfth International Conference on Learning Representations},
year={2024},
url={https://openreview.net/forum?id=xpw7V0P136}
}

@misc{Sarkar24,
      title={When {C}opilot Becomes Autopilot: {G}enerative {AI}'s Critical Risk to Knowledge Work and a Critical Solution}, 
      author={Advait Sarkar and Xiaotong and Xu and Neil Toronto and Ian Drosos and Christian Poelitz},
      year={2024},
      eprint={2412.15030},
      archivePrefix={arXiv},
      primaryClass={cs.HC},
      url={https://arxiv.org/abs/2412.15030}, 
}

@inproceedings{Eladawy24,
  author = {Hadeel Eladawy and Claire {Le Goues} and Yuriy Brun},
  title = {Automated Program Repair, What Is It Good For? {Not} Absolutely Nothing!},
  booktitle = {International Conference on Software Engineering},
  address = {Lisbon, Portugal},
  month = {April},
  date = {14--20},
  year = {2024},
  pages = {1017--1029},
  doi = {10.1145/3597503.3639095},   
}

@article{Candrian22,
title = {Rise of the machines: Delegating decisions to autonomous {AI}},
journal = {Computers in Human Behavior},
volume = {134},
pages = {107308},
year = {2022},
issn = {0747-5632},
doi = {https://doi.org/10.1016/j.chb.2022.107308},
url = {https://www.sciencedirect.com/science/article/pii/S0747563222001303},
author = {Cindy Candrian and Anne Scherer}}

@inproceedings{First23fse,
  author = {Emily First and Markus Rabe and Talia Ringer and Yuriy Brun},
  title = {Baldur: {Whole}-Proof Generation and Repair with Large Language Models},
  booktitle = {ACM Joint European Software Engineering Conference and
  Symposium on the Foundations of Software Engineering (ESEC/FSE)},
  pages = {1229--1241},
  month = {December},
  year = {2023},
  date = {6--8},
  address = {San Francisco, CA, USA},
  doi = {10.1145/3611643.3616243},
 }

@inproceedings{Motwani19icse,
  author = {Manish Motwani and Yuriy Brun},
  title = {Automatically Generating Precise Oracles from Structured Natural Language Specifications},
  booktitle = {International Conference on Software Engineering},
  pages = {188--199},
  address = {Montreal, QC, Canada},
  month = {May},
  date = {29--31},
  year = {2019},
  doi = {10.1109/ICSE.2019.00035},
}

@inproceedings{Goffi16,
  author = {Alberto Goffi and Alessandra Gorla and Michael D. Ernst and Mauro Pezz{\`{e}}},
  title = {Automatic generation of oracles for exceptional behaviors},
  booktitle = {International Symposium on Software Testing and Analysis (ISSTA)},
  pages = {213--224},
  address = {Saarbr{\"{u}}cken, Genmany},
  month = {July},
  year = {2016},
  doi = {10.1145/2931037.2931061},
}

@inproceedings{Blasi18,
author = {Blasi, Arianna and Goffi, Alberto and Kuznetsov, Konstantin and Gorla, Alessandra and Ernst, Michael D. and Pezz\`{e}, Mauro and Castellanos, Sergio Delgado},
title = {Translating code comments to procedure specifications},
year = {2018},
isbn = {9781450356992},
publisher = {Association for Computing Machinery},
address = {New York, NY, USA},
url = {https://doi.org/10.1145/3213846.3213872},
doi = {10.1145/3213846.3213872},
pages = {242–253},
numpages = {12},
location = {Amsterdam, Netherlands},
series = {ISSTA 2018}
}

@INPROCEEDINGS{Tan12,
  author={Tan, Shin Hwei and Marinov, Darko and Tan, Lin and Leavens, Gary T.},
  booktitle={2012 IEEE Fifth International Conference on Software Testing, Verification and Validation}, 
  title={@t{C}omment: {T}esting Javadoc Comments to Detect Comment-Code Inconsistencies}, 
  year={2012},
  volume={},
  number={},
  pages={260-269},
  doi={10.1109/ICST.2012.106}}

@INPROCEEDINGS{Pandita12,
  author={Pandita, Rahul and Xiao, Xusheng and Zhong, Hao and Xie, Tao and Oney, Stephen and Paradkar, Amit},
  booktitle={2012 34th International Conference on Software Engineering (ICSE)}, 
  title={Inferring method specifications from natural language {API} descriptions}, 
  year={2012},
  volume={},
  number={},
  pages={815-825},
  doi={10.1109/ICSE.2012.6227137}}

@INPROCEEDINGS{Zhong09,
  author={Zhong, Hao and Zhang, Lu and Xie, Tao and Mei, Hong},
  booktitle={International Conference on Automated Software Engineering}, 
  title={Inferring Resource Specifications from Natural Language {API} Documentation}, 
  year={2009},
  volume={},
  number={},
  pages={307-318},
  doi={10.1109/ASE.2009.94}}

@article{Tan07,
author = {Tan, Lin and Yuan, Ding and Krishna, Gopal and Zhou, Yuanyuan},
title = {/*icomment: bugs or bad comments?*/},
year = {2007},
issue_date = {December 2007},
publisher = {Association for Computing Machinery},
address = {New York, NY, USA},
volume = {41},
number = {6},
issn = {0163-5980},
url = {https://doi.org/10.1145/1323293.1294276},
doi = {10.1145/1323293.1294276},
journal = {SIGOPS Oper. Syst. Rev.},
month = oct,
pages = {145–158},
numpages = {14}
}

@INPROCEEDINGS{Zhou17,
  author={Zhou, Yu and Gu, Ruihang and Chen, Taolue and Huang, Zhiqiu and Panichella, Sebastiano and Gall, Harald},
  booktitle={International Conference on Software Engineering}, 
  title={Analyzing {API}s Documentation and Code to Detect Directive Defects}, 
  year={2017},
  volume={},
  number={},
  pages={27-37},
  doi={10.1109/ICSE.2017.11}}

@inproceedings{Ruan25,
author = {Ruan, Haifeng and Zhang, Yuntong and Roychoudhury, Abhik},
title = {{S}pec{R}over: {C}ode Intent Extraction via {LLMs}},
year = {2025},
isbn = {9798331505691},
publisher = {IEEE Press},
url = {https://doi.org/10.1109/ICSE55347.2025.00080},
doi = {10.1109/ICSE55347.2025.00080},
booktitle = {International Conference on Software Engineering},
pages = {963–974},
numpages = {12},
location = {Ottawa, Ontario, Canada},
series = {ICSE '25}
}

@article{Grandel26,
author = {Grandel, Skyler and Andersen, Scott Thomas and Huang, Yu and Leach, Kevin},
title = {{C}om{C}at: {E}xpertise-Guided Context Generation to Enhance Code Comprehension},
year = {2026},
issue_date = {March 2026},
publisher = {Association for Computing Machinery},
address = {New York, NY, USA},
volume = {35},
number = {3},
issn = {1049-331X},
url = {https://doi.org/10.1145/3742475},
doi = {10.1145/3742475},
month = feb,
articleno = {82},
numpages = {30}
}

@ARTICLE{Borstler16,
  author={B\"{o}rstler, J\"{u}rgen and Paech, Barbara},
  journal={IEEE Transactions on Software Engineering}, 
  title={The Role of Method Chains and Comments in Software Readability and Comprehension—An Experiment}, 
  year={2016},
  volume={42},
  number={9},
  pages={886-898},
  doi={10.1109/TSE.2016.2527791}}

@inproceedings{Stapleton20,
author = {Stapleton, Sean and Gambhir, Yashmeet and LeClair, Alexander and Eberhart, Zachary and Weimer, Westley and Leach, Kevin and Huang, Yu},
title = {A Human Study of Comprehension and Code Summarization},
year = {2020},
isbn = {9781450379588},
publisher = {Association for Computing Machinery},
address = {New York, NY, USA},
url = {https://doi.org/10.1145/3387904.3389258},
doi = {10.1145/3387904.3389258},
booktitle = {Proceedings of the 28th International Conference on Program Comprehension},
pages = {2–13},
numpages = {12},
location = {Seoul, Republic of Korea},
series = {ICPC '20}
}

@InProceedings{Khamis10,
author="Khamis, Ninus
and Witte, Ren{\'e}
and Rilling, Juergen",
editor="Hopfe, Christina J.
and Rezgui, Yacine
and M{\'e}tais, Elisabeth
and Preece, Alun
and Li, Haijiang",
title="Automatic Quality Assessment of Source Code Comments: {T}he {J}avadoc{M}iner",
booktitle="Natural Language Processing and Information Systems",
year="2010",
publisher="Springer Berlin Heidelberg",
address="Berlin, Heidelberg",
pages="68--79",
isbn="978-3-642-13881-2"
}

@article{Steidl13,
  title={Quality analysis of source code comments},
  author={Daniela Steidl and Benjamin Hummel and Elmar J{\"u}rgens},
  journal={International Conference on Program Comprehension (ICPC)},
  year={2013},
  pages={83-92},
  url={https://api.semanticscholar.org/CorpusID:16657129}
}

@INPROCEEDINGS{Abid15,
  author={Abid, Nahla J. and Dragan, Natalia and Collard, Michael L. and Maletic, Jonathan I.},
  booktitle={2015 IEEE International Conference on Software Maintenance and Evolution (ICSME)}, 
  title={Using stereotypes in the automatic generation of natural language summaries for {C}++ methods}, 
  year={2015},
  volume={},
  number={},
  pages={561-565},
  doi={10.1109/ICSM.2015.7332514}}

@InProceedings{Allamanis16,
  title = 	 {A Convolutional Attention Network for Extreme Summarization of Source Code},
  author = 	 {Allamanis, Miltiadis and Peng, Hao and Sutton, Charles},
  booktitle = 	 {International Conference on Machine Learning},
  pages = 	 {2091--2100},
  year = 	 {2016},
  editor = 	 {Balcan, Maria Florina and Weinberger, Kilian Q.},
  volume = 	 {48},
  series = 	 {Proceedings of Machine Learning Research},
  address = 	 {New York, New York, USA},
  month = 	 {20--22 Jun},
  publisher =    {PMLR},
  url = 	 {https://proceedings.mlr.press/v48/allamanis16.html}
}

@INPROCEEDINGS{Eddy13,
  author={Eddy, Brian P. and Robinson, Jeffrey A. and Kraft, Nicholas A. and Carver, Jeffrey C.},
  booktitle={International Conference on Program Comprehension}, 
  title={Evaluating source code summarization techniques: {R}eplication and expansion}, 
  year={2013},
  volume={},
  number={},
  pages={13-22},
  doi={10.1109/ICPC.2013.6613829}}

@inproceedings{Feng20,
    title = "{C}ode{BERT}: {A} Pre-Trained Model for Programming and Natural Languages",
    author = "Feng, Zhangyin  and
      Guo, Daya  and
      Tang, Duyu  and
      Duan, Nan  and
      Feng, Xiaocheng  and
      Gong, Ming  and
      Shou, Linjun  and
      Qin, Bing  and
      Liu, Ting  and
      Jiang, Daxin  and
      Zhou, Ming",
    editor = "Cohn, Trevor  and
      He, Yulan  and
      Liu, Yang",
    booktitle = "Findings of the Association for Computational Linguistics: EMNLP 2020",
    month = nov,
    year = "2020",
    address = "Online",
    publisher = "Association for Computational Linguistics",
    url = "https://aclanthology.org/2020.findings-emnlp.139/",
    doi = "10.18653/v1/2020.findings-emnlp.139",
    pages = "1536--1547"
}

@article{Gao23,
author = {Gao, Shuzheng and Gao, Cuiyun and He, Yulan and Zeng, Jichuan and Nie, Lunyiu and Xia, Xin and Lyu, Michael},
title = {Code Structure–Guided Transformer for Source Code Summarization},
year = {2023},
issue_date = {January 2023},
publisher = {Association for Computing Machinery},
address = {New York, NY, USA},
volume = {32},
number = {1},
issn = {1049-331X},
url = {https://doi.org/10.1145/3522674},
doi = {10.1145/3522674},
journal = {ACM Trans. Softw. Eng. Methodol.},
month = feb,
articleno = {23},
numpages = {32}
}

@inproceedings{Haiduc10,
author = {Haiduc, Sonia and Aponte, Jairo and Marcus, Andrian},
title = {Supporting program comprehension with source code summarization},
year = {2010},
isbn = {9781605587196},
publisher = {Association for Computing Machinery},
address = {New York, NY, USA},
url = {https://doi.org/10.1145/1810295.1810335},
doi = {10.1145/1810295.1810335},
booktitle = {32nd International Conference on Software Engineering - Volume 2},
pages = {223–226},
numpages = {4},
location = {Cape Town, South Africa},
series = {ICSE '10}
}

@inproceedings{Iyer16,
    title = "Summarizing Source Code using a Neural Attention Model",
    author = "Iyer, Srinivasan  and
      Konstas, Ioannis  and
      Cheung, Alvin  and
      Zettlemoyer, Luke",
    editor = "Erk, Katrin  and
      Smith, Noah A.",
    booktitle = "Proceedings of the 54th Annual Meeting of the Association for Computational Linguistics (Volume 1: Long Papers)",
    month = aug,
    year = "2016",
    address = "Berlin, Germany",
    publisher = "Association for Computational Linguistics",
    url = "https://aclanthology.org/P16-1195/",
    doi = "10.18653/v1/P16-1195",
    pages = "2073--2083"
}

@inproceedings{Khan23,
author = {Khan, Junaed Younus and Uddin, Gias},
title = {Automatic Code Documentation Generation Using {GPT}-3},
year = {2023},
isbn = {9781450394758},
publisher = {Association for Computing Machinery},
address = {New York, NY, USA},
url = {https://doi.org/10.1145/3551349.3559548},
doi = {10.1145/3551349.3559548},
booktitle = {International Conference on Automated Software Engineering},
articleno = {174},
numpages = {6},
location = {Rochester, MI, USA},
series = {ASE '22}
}

@misc{Liu21,
      title={Retrieval-Augmented Generation for Code Summarization via Hybrid GNN}, 
      author={Shangqing Liu and Yu Chen and Xiaofei Xie and Jingkai Siow and Yang Liu},
      year={2021},
      eprint={2006.05405},
      archivePrefix={arXiv},
      primaryClass={cs.LG},
      url={https://arxiv.org/abs/2006.05405}, 
}

@inproceedings{McBurney14,
author = {McBurney, Paul W. and McMillan, Collin},
title = {Automatic documentation generation via source code summarization of method context},
year = {2014},
isbn = {9781450328791},
publisher = {ACM},
address = {New York, NY, USA},
url = {https://doi.org/10.1145/2597008.2597149},
doi = {10.1145/2597008.2597149},
booktitle = {Proceedings of the 22nd International Conference on Program Comprehension},
pages = {279–290},
numpages = {12},
location = {Hyderabad, India},
series = {ICPC 2014}
}

@inproceedings{Parvez21,
    title = "Retrieval Augmented Code Generation and Summarization",
    author = "Parvez, Md Rizwan  and
      Ahmad, Wasi  and
      Chakraborty, Saikat  and
      Ray, Baishakhi  and
      Chang, Kai-Wei",
    editor = "Moens, Marie-Francine  and
      Huang, Xuanjing  and
      Specia, Lucia  and
      Yih, Scott Wen-tau",
    booktitle = "Findings of the Association for Computational Linguistics: EMNLP 2021",
    month = nov,
    year = "2021",
    address = "Punta Cana, Dominican Republic",
    publisher = "Association for Computational Linguistics",
    url = "https://aclanthology.org/2021.findings-emnlp.232/",
    doi = "10.18653/v1/2021.findings-emnlp.232",
    pages = "2719--2734"}

@inproceedings{Phan21,
    title = "{C}o{T}ex{T}: {M}ulti-task Learning with Code-Text Transformer",
    author = "Phan, Long  and
      Tran, Hieu  and
      Le, Daniel  and
      Nguyen, Hieu  and
      Annibal, James  and
      Peltekian, Alec  and
      Ye, Yanfang",
    editor = "Lachmy, Royi  and
      Yao, Ziyu  and
      Durrett, Greg  and
      Gligoric, Milos  and
      Li, Junyi Jessy  and
      Mooney, Ray  and
      Neubig, Graham  and
      Su, Yu  and
      Sun, Huan  and
      Tsarfaty, Reut",
    booktitle = "Proceedings of the 1st Workshop on Natural Language Processing for Programming (NLP4Prog 2021)",
    month = aug,
    year = "2021",
    address = "Online",
    publisher = "Association for Computational Linguistics",
    url = "https://aclanthology.org/2021.nlp4prog-1.5/",
    doi = "10.18653/v1/2021.nlp4prog-1.5",
    pages = "40--47",
}

@inproceedings{Sridhara10,
author = {Sridhara, Giriprasad and Hill, Emily and Muppaneni, Divya and Pollock, Lori and Vijay-Shanker, K.},
title = {Towards automatically generating summary comments for {Java} methods},
year = {2010},
isbn = {9781450301169},
publisher = {ACM},
address = {New York, NY, USA},
url = {https://doi.org/10.1145/1858996.1859006},
doi = {10.1145/1858996.1859006},
booktitle = {International Conference on Automated Software Engineering},
pages = {43–52},
numpages = {10},
location = {Antwerp, Belgium},
series = {ASE '10}
}

@inproceedings{Wan18,
author = {Wan, Yao and Zhao, Zhou and Yang, Min and Xu, Guandong and Ying, Haochao and Wu, Jian and Yu, Philip S.},
title = {Improving automatic source code summarization via deep reinforcement learning},
year = {2018},
isbn = {9781450359375},
publisher = {Association for Computing Machinery},
address = {New York, NY, USA},
url = {https://doi.org/10.1145/3238147.3238206},
doi = {10.1145/3238147.3238206},
booktitle = { International Conference on Automated Software Engineering},
pages = {397–407},
numpages = {11},
location = {Montpellier, France},
series = {ASE '18}
}

@inproceedings{Wang21,
    title = "{C}ode{T}5: {I}dentifier-aware Unified Pre-trained Encoder-Decoder Models for Code Understanding and Generation",
    author = "Wang, Yue  and
      Wang, Weishi  and
      Joty, Shafiq  and
      Hoi, Steven C.H.",
    editor = "Moens, Marie-Francine  and
      Huang, Xuanjing  and
      Specia, Lucia  and
      Yih, Scott Wen-tau",
    booktitle = "Conference on Empirical Methods in Natural Language Processing",
    month = nov,
    year = "2021",
    address = "Online and Punta Cana, Dominican Republic",
    publisher = "Association for Computational Linguistics",
    url = "https://aclanthology.org/2021.emnlp-main.685/",
    doi = "10.18653/v1/2021.emnlp-main.685",
    pages = "8696--8708",
}

@inproceedings{Lin04,
  title={{ROUGE}: {A} Package for Automatic Evaluation of Summaries},
  author={Chin-Yew Lin},
  booktitle={Annual Meeting of the Association for Computational Linguistics},
  year={2004},
  url={https://api.semanticscholar.org/CorpusID:964287}
}

@article{Li26,
title = {Do comments and expertise still matter? An experiment on programmers' adoption of {AI}-generated {JavaScript} code},
journal = {Journal of Systems and Software},
volume = {231},
pages = {112634},
year = {2026},
issn = {0164-1212},
doi = {https://doi.org/10.1016/j.jss.2025.112634},
url = {https://www.sciencedirect.com/science/article/pii/S0164121225003036},
author = {Changwen Li and Christoph Treude and Ofir Turel}
}

@article{West22,
author = {Robert M West},
title ={Best practice in statistics: {T}he use of log transformation},

journal = {Annals of Clinical Biochemistry},
volume = {59},
number = {3},
pages = {162-165},
year = {2022},
doi = {10.1177/00045632211050531},
note ={PMID: 34666549},
URL = {     
        https://doi.org/10.1177/00045632211050531
},
eprint = {     
        https://doi.org/10.1177/00045632211050531
}
}

@article{rothman1990no,
  title={No adjustments are needed for multiple comparisons},
  author={Rothman, Kenneth J},
  journal={Epidemiology},
  volume={1},
  number={1},
  pages={43--46},
  year={1990},
  publisher={LWW}
}

@article{hsieh2005three,
  title={Three approaches to qualitative content analysis},
  author={Hsieh, Hsiu-Fang and Shannon, Sarah E},
  journal={Qualitative health research},
  volume={15},
  number={9},
  pages={1277--1288},
  year={2005},
  publisher={Sage Publications Sage CA: Thousand Oaks, CA}
}

@article{Curran18,
author = {Curran-Everett, Douglas},
title = {Explorations in statistics: the log transformation},
journal = {Advances in Physiology Education},
volume = {42},
number = {2},
pages = {343-347},
year = {2018},
doi = {10.1152/advan.00018.2018},
note ={PMID: 29761718},
URL = {    
        https://doi.org/10.1152/advan.00018.2018
},
eprint = { 
        https://doi.org/10.1152/advan.00018.2018}}

@article{Hayet25,
author = {Hayet, Ishrak and Scott, Adam and d'Amorim, Marcelo},
title = {{ChatAssert}: {LLM}-Based Test Oracle Generation With External Tools Assistance},
year = {2025},
issue_date = {Jan. 2025},
publisher = {IEEE Press},
volume = {51},
number = {1},
issn = {0098-5589},
url = {https://doi.org/10.1109/TSE.2024.3519159},
doi = {10.1109/TSE.2024.3519159},
journal = {IEEE Trans. Softw. Eng.},
month = jan,
pages = {305–319},
numpages = {15}
}

@INPROCEEDINGS{Li25Extracting,
  author={Li, Hui and Dong, Zhen and Wang, Siao and Zhang, Hui and Shen, Liwei and Peng, Xin and She, Dongdong},
  booktitle={International Conference on Program Comprehension (ICPC)}, 
  title={Extracting Formal Specifications From Documents Using {LLMS} for Test Automation}, 
  year={2025},
  volume={},
  number={},
  pages={1-12},
  doi={10.1109/ICPC66645.2025.00039}}

@INPROCEEDINGS{Molinelli25,
  author={Molinelli, Davide and Grazia, Luca Di and Martin-Lopez, Alberto and Ernst, Michael D. and Pezzè, Mauro},
  booktitle={International Conference on Automated Software Engineering (ASE)}, 
  title={Do {LLMs} Generate Useful Test Oracles? An Empirical Study with an Unbiased Dataset}, 
  year={2025},
  volume={},
  number={},
  pages={278-290},
  doi={10.1109/ASE63991.2025.00031}}

@misc{Wu26,
      title={Viverra: Text-to-Code with Guarantees}, 
      author={Haoze Wu and Rocky Klopfenstein and Keith Farkas and Nina Narodytska},
      year={2026},
      eprint={2605.14972},
      archivePrefix={arXiv},
      primaryClass={cs.SE},
      url={https://arxiv.org/abs/2605.14972}, 
}

@inproceedings{Jahic24,
  author={Jasmin Jahi{\'{c}} and Ashkan Sami},
  booktitle={International Conference on Software Architecture (ICSA)}, 
  title={State of Practice: {LLMs} in Software Engineering and Software Architecture}, 
  year={2024},
  pages={311--318},
  doi={10.1109/ICSA-C63560.2024.00059},
}

@misc{StackOverflow2025Survey,
  title        = {2025 Stack Overflow Developer Survey},
  author       = {{Stack Overflow}},
  year         = {2025},
  url          = {https://survey.stackoverflow.co/2025},
  note         = {Accessed June 2026}
}

@inproceedings{bacchelli2013expectations,
  title={Expectations, outcomes, and challenges of modern code review},
  author={Bacchelli, Alberto and Bird, Christian},
  booktitle={2013 35th international conference on software engineering (ICSE)},
  pages={712--721},
  year={2013},
  organization={IEEE}
}

@article{zhong2026human,
  title={Human--{AI} Synergy in Agentic Code Review},
  author={Zhong, Suzhen and Noei, Shayan and Zou, Ying and Adams, Bram},
  journal={arXiv preprint arXiv:2603.15911},
  year={2026}
}

@article{pimenova2025good,
  title={Good vibrations? {A} qualitative study of co-creation, communication, flow, and trust in vibe coding},
  author={Pimenova, Veronica and Fakhoury, Sarah and Bird, Christian and Storey, Margaret-Anne and Endres, Madeline},
  journal={arXiv preprint arXiv:2509.12491},
  year={2025}
}

@article{baum2019associating,
  title={Associating working memory capacity and code change ordering with code review performance},
  author={Baum, Tobias and Schneider, Kurt and Bacchelli, Alberto},
  journal={Empirical Software Engineering},
  volume={24},
  number={4},
  pages={1762--1798},
  year={2019},
  publisher={Springer}
}

@inproceedings{lemieux2023codamosa,
  title={Codamosa: Escaping coverage plateaus in test generation with pre-trained large language models},
  author={Lemieux, Caroline and Inala, Jeevana Priya and Lahiri, Shuvendu K and Sen, Siddhartha},
  booktitle={International Conference on Software Engineering (ICSE)},
  pages={919--931},
  year={2023},
  organization={IEEE}
}

@article{zietsman2026specification,
  title={The Specification as Quality Gate: {Th}ree Hypotheses on {AI}-Assisted Code Review},
  author={Zietsman, Christo},
  journal={arXiv preprint arXiv:2603.25773},
  year={2026}
}

@article{lahiri2026intent,
  title={{Intent formalization: A grand} challenge for reliable coding in the age of {AI} agents},
  author={Lahiri, Shuvendu K},
  journal={arXiv preprint arXiv:2603.17150},
  year={2026}
}

@article{hou2024large,
  title={Large language models for software engineering: {A} systematic literature review},
  author={Hou, Xinyi and Zhao, Yanjie and Liu, Yue and Yang, Zhou and Wang, Kailong and Li, Li and Luo, Xiapu and Lo, David and Grundy, John and Wang, Haoyu},
  journal={ACM Transactions on Software Engineering and Methodology},
  volume={33},
  number={8},
  pages={1--79},
  year={2024},
  publisher={ACM New York, NY}
}

@InProceedings{prasad_et_al:LIPIcs.ECOOP.2026.22,
  author =	{Prasad, Siddhartha and Austen, Skyler and Fisler, Kathi and Krishnamurthi, Shriram},
  title =	{{Meaningful Human-in-the-Loop Checking of Gen{AI} Synthesis for Restricted Languages}},
  booktitle =	{40th European Conference on Object-Oriented Programming (ECOOP 2026)},
  pages =	{22:1--22:31},
  year =	{2026},
  volume =	{372},
  URL =		{https://drops.dagstuhl.de/entities/document/10.4230/LIPIcs.ECOOP.2026.22},
  URN =		{urn:nbn:de:0030-drops-261181},
  doi =		{10.4230/LIPIcs.ECOOP.2026.22}
}

\end{document}